\documentclass{aa}  

\usepackage{color}
\usepackage{graphicx}
\usepackage{txfonts}
\usepackage{natbib}

\begin{document} 
\date{Accepted 22/06/2021}

   \title{Probing protoplanetary disk evolution in the Chamaeleon II region}

   \author{M. Villenave
          \inst{1,}\inst{2,}\inst{3}
          \and
          F. M\'enard\inst{1}
          \and
          W. R. F. Dent\inst{4}
          \and
          M. Benisty\inst{5,}\inst{1}
          \and
          G. van der Plas\inst{1}
         \and 
         J. P. Williams\inst{6}
          \and
          M. Ansdell\inst{7}
          \and
          \'A. Ribas\inst{2}
 	      \and 
         C. Caceres\inst{8,}\inst{9}
		  \and 
         H.~Canovas\inst{10}
         \and 
         L. Cieza\inst{11}
 		\and
         A. Hales\inst{4,}\inst{12}
         \and 
         I. Kamp\inst{13}
         \and
         C. Pinte\inst{14, }\inst{1}
         \and 
         D. A. Principe\inst{15}
         \and 
         M. R. Schreiber\inst{9,}\inst{16}
    }

   \institute{Univ. Grenoble Alpes, CNRS, IPAG, 38000 Grenoble, France\\
   			\email{marion.f.villenave@jpl.nasa.gov}
              \and 
	  	European Southern Observatory, Alonso de C\'ordova 3107, Vitacura, Casilla 19001, Santiago 19, Chile 
	  	    \and
	  	    Jet Propulsion Laboratory, California Institute of Technology, 4800 Oak Grove Drive, Pasadena, CA 91109, USA
             \and
    		Joint ALMA Observatory, Alonso de C\'ordova 3107, Vitacura 763-0355, Santiago, Chile
              \and 
              Unidad Mixta Internacional Franco-Chilena de Astronom\'ia (CNRS, UMI 3386), Departamento de Astronom\'ia, Universidad de
            Chile, Camino El Observatorio 1515, Las Condes, Santiago, Chile
            \and 
            Institute for Astronomy, University of Hawaii, Honolulu, Hawaii, USA
            \and
            NASA Headquarters, 300 E Street SW, Washington, DC 20546, USA
            \and 
            Departamento de Ciencias Fisicas, Facultad de Ciencias Exactas, Universidad Andres Bello. Av. Fernandez Concha 700, Las Condes, Santiago, Chile
            \and 
            N\'ucleo Milenio de Formaci\'on Planetaria (NPF), Av. Gran Breta\~na 1111, Valparaiso, Chile
            \and 
            Aurora Technology for ESA/ESAC, Camino bajo del Castillo s/n, Urbanizaci\'on Villafranca del Castillo, Villanueva de la Ca\~nada, 28692 Madrid, Spain
            \and 
            N\'ucleo de Astronom\'ia,  Facultad de Ingenier{\'i}a y Ciencias, Universidad Diego Portales,  Av. Ejercito 441, Santiago, Chile
		\and 
            National Radio Astronomy Observatory, 520 Edgemont Road, Charlottesville, Virginia, 22903-2475, United States
            \and
            Kapteyn Astronomical Institute, University of Groningen, Landleven 12, NL-9747 AD Groningen, The Netherlands
            \and
            School of Physics and Astronomy, Monash University, Clayton Vic 3800, Australia
            \and 
            Massachusetts Institute of Technology, Kavli Institute for Astrophysics and Space Research, Cambridge, MA 02109, USA
	\and
            Universidad T\'ecnica Federico Santa Mar\'ia, Departamento de F\'isica, Avenida Espa\~na 1680, Valpara\'iso, Chile
		}
        
 \abstract
 {Characterizing the evolution of protoplanetary disks is necessary to improve our understanding of planet formation. Constraints on both dust and gas are needed to determine the dominant disk dissipation mechanisms.} 
 {We aim to compare the disk dust masses in the Chamaeleon~II (Cha~II) star-forming region with other regions with ages between 1 and 10\,Myr.} 
 {We use ALMA band 6 observations (1.3 mm) to survey 29 protoplanetary disks in Cha~II. Dust mass estimates are derived from the continuum data.} 
 {Out of our initial sample of 29 disks, we detect 22 sources in the continuum, 10 in $^{12}$CO, 3 in $^{13}$CO, and none in C$^{18}$O (J=2-1).  Additionally, we detect two companion candidates in the continuum and $^{12}$CO emission. Most disk dust masses are lower than~10~M$_{\oplus}$, assuming thermal emission from optically thin dust. Including non-detections, we derive a median dust mass of $4.5 \pm 1.5$~M$_\oplus$ from survival analysis. 
We compare consistent estimations of the distributions of the disk dust mass and the disk-to-stellar mass ratios in Cha~II with six other low mass and isolated star-forming regions in the age range of 1-10~Myr: Upper Sco, CrA, IC~348, Cha~I, Lupus, and Taurus. When comparing the dust-to-stellar mass ratio, we find that the masses of disks in Cha~II are statistically different from those in Upper Sco and Taurus, and we confirm that disks in Upper Sco, the oldest region of the sample, are statistically less massive than in all other regions. Performing a second statistical test of the dust mass distributions from similar mass bins, we find no statistical differences between these regions and Cha~II.}
{We interpret these trends, most simply, as a sign of decline in the disk dust masses with time or dust evolution. Different global initial conditions in star-forming regions may also play a role, but their impact on the properties of a disk population is difficult to isolate in star-forming regions lacking nearby massive stars.} 
{}

   \maketitle

\section{Introduction}

Planets are thought to form in gas- and dust-rich protoplanetary disks that orbit young stars. Characterizing the physical properties and evolutionary mechanisms of protoplanetary disks is therefore essential for the understanding of planet formation. Various infrared studies have shown that the typical dissipation timescale of protoplanetary disks is about 3~Myr~\citep[e.g.,][]{Mamajek_2009, ribas_2014}, the oldest disks typically being up to 10~Myr old. However, these observations trace either small, warm dust particles (in the continuum) or accretion signatures~\citep{fedele_2010}, which are not sensitive to the dissipation of the bulk gas and dust mass, more relevant for planet formation.

More recently, the Atacama Large Millimeter/submillimeter Array (ALMA) has allowed to perform a number of statistical surveys of the disk dust mass to be performed in various star-forming regions. Many surveys have focused on relatively young, $1-3$ Myr, regions~\citep[Taurus, Chamaeleon~I, Lupus, Ophiuchus, IC348, ONC, CrA, OMC-2, Lynds 1641;][]{Andrews_2013, Pascucci_2016, Ansdell_2016, Cieza_2018, Ruiz-Rodriguez_2018, Eisner_2018, Cazzoletti_2019, van_Terwisga_2019, Grant_2021}. They showed that the disk dust masses estimated in these young regions are in general larger than those measured in a more evolved star-forming region~\citep[$5-10$\,Myr, Upper Sco;][]{Barenfeld_2016, VanDerPlas_2016}. This result indicates that the disk dust mass decreases with time. 
In addition, several surveys of intermediate age star-forming regions have also been performed~\citep[$\sigma$-Orionis, $\lambda$-Orionis;][]{Ansdell_2017, Ansdell_2020}. However, those mostly focus on regions that include massive stars. They find that massive stars can have an important impact on the evolution of disks.
Further studies of intermediately aged regions, unaffected by external factors of disk evolution, are required to study dust disk dissipation.

In this context, the Chamaeleon II (Cha~II) star-forming region is of particular interest. Its age was historically estimated to be $4\pm 2$~Myr~\citep{Spezzi_2008}, which made it a good choice for studying the evolution of gas and dust content. However, we note that a recent study revised the age of the region using \emph{Gaia Data Release 2} (DR2) distances~\citep{Galli_2020b}. That study suggests that Cha~II is significantly younger, with a median age of $1-2$ Myr. This new age is consistent with the high disk fraction observed in previous infrared surveys~\citep[e.g.,][]{ribas_2014}. Cha~II is a close-by region, located at an average distance of 198~pc~\citep{Dzib_2018, Galli_2020b}. It has been the target of several infrared studies~\citep{Alcala_2008, Spezzi_2008, Spezzi_2013}, which have shown that the region is quite isolated and does not contain high mass stars. Thus, the evolution of its disks is likely not driven by external factors, permitting the study of isolated disk evolution.

In this paper, we present an ALMA survey of the Class II disks of Cha~II. We describe our ALMA observations and data reduction in Sect.~\ref{sec:observations}, and present the continuum and CO line measurements in Sect.~\ref{sec:results}. In Sect.~\ref{sec:prop_dust}, we analyze the dust properties of the Cha~II disks. We estimate the dust masses and compare the dust-to-star mass ratio with different star-forming regions. 
Our results are summarized in Sect.~\ref{summary}.

\begin{table*}[h!]
\caption{Stellar parameters.}
\centering
\begin{tabular}{lcllcccc}  
\hline\hline
Source  & RA (h.m.s) & Dec (deg.m.s)& Cloud  & SpT$^{(a)}$ & Class$^{(a)}$ &d (pc)&  M$_{\star }$  (M$_{\odot }$)$^{(e)}$\\
\hline
J13022287-7734494& 13 02 22.8 & -77 34 49.6 &Member & M5 &II &200.3 $\pm$ 2.1$^{(b)}$ & 0.19 {\small (0.13, 0.26)}\\
J13071806-7740529 &13 07 18.0 & -77 40 53.0$^{(a)}$& Member & M4.5 &II &197.9 $\pm$ 3.1 $^{(b)}$ & 0.23 {\small (0.17, 0.31)}\\
J13082714-7743232 &13 08 27.2 &-77 43 23.4$^{(a)}$& Member &M4.5 &II &199.5 $\pm$ 2.6 $^{(b)}$ & 0.22 {\small (0.16, 0.31)} \\
CM Cha & 13 02  13.4& -76 37 57.9& Member &K7 &II &194.1 $\pm$1 .5 $^{(b)}$ & 0.81 {\small (0.63, 0.98)}\\
Hn 22 &13 04   22.8& -76 50 05.6 & Member & M2&II &197.1 $\pm$ 1.9 $^{(b)}$& 0.44 {\small (0.34, 0.56)}\\
Hn 23 &13 04   24.0& -76 50 01.3 & Member &K5 & II &198.0 $\pm$ 1.8 $^{(b)}$& 1.06 {\small (0.88, 1.21)}\\
Hn 24 A&  13 04   55.6& -77 39   49.3 & Member & M0&II &196.7 $\pm$ 1.9 $^{(b)}$& 0.57 {\small (0.48, 0.72)}\\
Hn 24 B & 13 04 55.6 & -77 39 51.0& Uncertain& ...& ...&... &... \\
Hn 25 &  13 05    08.4& -77 33   42.8 & Member &M2.5 & II &198.0 $\pm$ 2.6 $^{(b)}$ & 0.37 {\small (0.27, 0.49)}\\
Hn 26 &13 07 48.5& -77 41 21.7$^{(a)}$& Member & M2& II&197.5$\pm$ 2.5 $^{(b)}$& 0.44 {\small (0.33, 0.56)}\\
IRAS12496-7650& 12 53   17.1& -77 07   10.9 & Background &F0 & II&243.7 $\pm$ 22.0$^{(c)}$&>1.4\\
IRAS12500-7658 &12 53   42.7& -77 15   11.8 &Uncertain & K5 & I & ... & 0.08 {\small (0.06, 0.11)}\\
IRAS12535-7623 & 12 57   11.6 & -76 40   11.7 & Likely & M0&II &180.0 $\pm$ 9.5$^{(d)}$ & 0.58 {\small (0.49, 0.72)}\\
ISO-CHAII 13 & 12 58 06.7 & -77 09 09.2$^{(a)}$&Uncertain &M7&II &... & 0.08 {\small (0.06, 0.11)}\\
J130059.3-771403 &13 00   59.2& -77 14    03.0 &Likely &K3 & II &200.2 $\pm$ 6.0$^{(d)}$&...\\
J130521.7-773810 &13 05 21.7 &-77 38 10.1$^{(a)}$& Likely &...& F &192.2 $\pm$ 3.6$^{(d)}$&...\\
J130529.0-774140 &13 05 27.1& -77 41 21.5$^{(a)}$&Uncertain  &... & II &... &...\\
Sz\,46 &  12 56   33.5& -76 45   45.6& Member  &M1 &II &195.6 $\pm$ 2.3$^{(b)}$ & 0.54 {\small (0.42, 0.66)}\\
Sz\,49 &13 00   53.1& -76 54   15.3 & Member &M0.5 &II &194.3 $\pm$ 1.8$^{(b)}$ & 0.57 {\small (0.46, 0.70)}\\
Sz\,50 &13 00   55.2& -77 10   22.4 & Foreground & M3 &II &147.4 $\pm$ 13.0$^{(c)} $ & 0.33 {\small (0.29, 0.40)}\\
Sz\,51 &  13 01   58.8& -77 51   21.9  & Member &K8.5 &II &195.9 $\pm$ 1.7$^{(b)}$ & 0.71 {\small (0.58, 0.86)} \\
Sz\,52 &13 04   24.8& -77 52   30.4 & Member &M2.5 &II &199.4 $\pm$ 2.4$^{(b)}$ & 0.40 {\small (0.31, 0.51)}\\
Sz\,53 &13 05   12.6& -77 30   52.7 & Member &M1 &II &198.3 $\pm$ 2.4 $^{(b)}$ & 0.54 {\small (0.42, 0.66)}\\
Sz\,54 &13 05   20.6& -77 39    01.6& Member & K5&II &196.7 $\pm$ 1.7$^{(b)}$& 1.03 {\small (0.84, 1.24)} \\
Sz\,56 &13 06   38.7& -77 30   35.4& Member & M4&II &194.1 $\pm$ 2.1 $^{(b)}$& 0.25 {\small (0.21, 0.32)}\\
Sz\,58 &13 06   57.3& -77 23   41.7 & Member & K5 &II &193.5 $\pm$ 1.7$^{(b)}$ & 1.01 {\small (0.87, 1.14)} \\
Sz\,59 A & 13 07 09.2& -77 30   30.9  & Likely & K7&II &193.3 $\pm$ 4.3$^{(d)}$ & 0.78 {\small (0.61, 0.99)}\\
Sz\,59 B & 13 07 09.1 & -77 30 30.2& Uncertain&... &... & ...&...\\
Sz\,61  &13 08 06.2& -77 55  05.4 & Member &K5 &II &199.0 $\pm$ 2.1$^{(b)}$ &  1.07 {\small (0.86, 1.26)}\\
Sz\,62 & 	13 09 50.4 & -77 57 23.9$^{(a)}$& Likely & M2.5&II &212.7 $\pm$ 5.8$^{(d)}$ & 0.36 {\small (0.27, 0.48)}\\
Sz\,63 & 13 10  04.1& -77 10 45.0~& Member & M3& II &200.0 $\pm$ 1.8$^{(b)}$ & 0.33 {\small (0.24, 0.45)}\\
\hline
 Sz\,48\,NE & 13 00 53.2 &  -77 09 09.2$^{(a)}$ & Unobserved &  M0.5 & II &198.6 $\pm$ 2.6$^{(b)}$  & 0.59 {\small (0.46, 0.71)}\\
 Sz\,48\,SW & 13 00 53.6 & -77 09 08.3$^{(a)}$ & Unobserved & M1 & II &197.8$\pm$ 2.6$^{(b)}$ & 0.54 {\small (0.43, 0.66)}\\
Sz\,55 & 13  06 30.5  & -77 34 00.1$^{(a)}$ & Unobserved & M2 & II &198.2 $\pm$ 2.5$^{(b)}$& 0.44 {\small (0.34, 0.57)}\\
 Sz\,57 & 13 06 56.6 & -77 23 09.5$^{(a)}$&  Unobserved& M5&  II &195.6 $\pm$ 3.1$^{(b)}$ & 0.23 {\small (0.20, 0.27)}\\
Sz\,60\,E &13  07 23.3 & -77 37 23.2$^{(a)}$& Unobserved & M4 & II &197.1$\pm$ 2.6$^{(b)}$ & 0.24 {\small (0.19, 0.33)}\\
 IRASF 13052-1653N & 13 09 11.0 & -77 09 44.1$^{(a)}$& Unobserved& M1.5 &  II &201.5 $\pm$ 1.5$^{(b)}$& 0.48 {\small (0.37, 0.61)}\\
IRASF 13052-7653NW &  13  09 09.8&   -77 09 43.5$^{(a)}$ & Unobserved & M0.5 & II &196.9 $\pm$1.8$^{(b)}$& 0.57 {\small (0.46, 0.71)} \\
Sz 64 & 13 14 03.8  & -77 53 07.8$^{(a)}$  & Unobserved & M5 & II & 196.5$\pm$ 2.6$^{(b)}$ & 0.20 {\small (0.13, 0.27)}\\
\hline
\end{tabular}
\tablefoot{Coordinates are either from \citet[][marked with $^{(a)}$]{Spezzi_2013} or, for the sources detected with ALMA (see Table~\ref{tab:cont-emission}), from our continuum fit presented in Section~\ref{sec:observation_cont}. The column "Cloud" refers to the cluster membership discussed in Sect.~\ref{sec:sample}. Stellar masses and confidence intervals are calculated with the pre-main sequence tracks by~\citet{baraffe_2015}, using luminosities and temperatures from~\citet{Spezzi_2008} rescaled using individual distances (Sect.~\ref{sec:methods}).}
\tablebib{ $^{(a)}$~\citet{Spezzi_2013}, $^{(b)}$~\citet{Galli_2020b}, $^{(c)}$~\citet{Gaia_DR3_2020}, $^{(d)}$~\citet{gaia_DR2_2018}, $^{(e)}$~This work.}
\label{tab:physical_param}
\end{table*}

\section{Observations and data reduction}
\label{sec:observations}

\subsection{Sample}
\label{sec:sample}
Our observations focus on 29 protoplanetary disks that were selected based on their infrared emission at 70\,$\mu$m from \emph{Herschel} observations~\citep{Spezzi_2013}. Specifically, we observed 18 out of the 19 Class~II disks detected at 70\,$\mu$m by \citet[][]{Spezzi_2013}, and 9 out of the 19 non-detected Class~II disks at 70\,$\mu$m (see Appendix~\ref{apdx:comparison}). We also observed one Class~I disk detected at 70\,$\mu$m and one "flat spectrum" source that was not detected at 70\,$\mu$m. To summarize, the sample includes one Class~I source, one flat spectrum source, and 27 Class II sources; two secondary sources~(around Hn\,24 and Sz\,59) were also detected in our ALMA observations (see Sect.~\ref{sec:observation_cont}), leading to a total sample of 31 objects. 

We checked the membership of each observed target by using distances from \emph{Gaia} data releases and the recent membership analysis performed by \citet[][their Table A.2.]{Galli_2020b}.  We find that 19 disks included in our sample were confirmed as members by \citet{Galli_2020b}. We classify these sources as "Members" in Table~\ref{tab:physical_param} and we report the individual distances as estimated by \citet{Galli_2020b}. We also identified 5 sources, rejected or not included in the study of \citet[][]{Galli_2020b}, as likely cluster members given their latest \emph{Gaia EDR3} distance~\citep{Gaia_DR3_2020}. Their \emph{Gaia EDR3} distances lie less than 20\,pc away from the average cluster distance. 
We classify these objects as "Likely" members in Table~\ref{tab:physical_param}, and report the individual distances calculated from the \emph{Gaia EDR3} catalog~\citep{Gaia_DR3_2020}.
Two disks, namely Sz\,50 and IRAS12496-7650, are located more than 40\,pc away from the average cluster distance, so they are possibly foreground and background objects, respectively. For these sources, we report the distances estimated using \emph{Gaia DR2} parallaxes in Table~\ref{tab:physical_param}, because the latest release produced less reliable results (larger parallax error for IRAS12496-7650 and no \emph{EDR3} measurement for Sz\,50). 
Finally, three disks (J130529.0-774140, IRAS12500-7658, ISO-CHAII 13) do not have a good (or any) parallax measurement in either \emph{Gaia} data releases, making them "Uncertain" cluster members.

To summarize, our sample is composed of 29 disks, including 19 Class~II confirmed members of Cha~II, 4 Class~II likely members, and 1 flat spectrum likely member. Two Class~II sources and one Class~I are uncertain members, and two Class~II are likely foreground and background systems, external to the Cha~II star-forming region. 
We report the membership of each source along with its adopted distance and stellar parameters in Table~\ref{tab:physical_param}. Additionally, 8 Class II sources, not observed in this study and observed but not detected at 70\,$\mu$m~\citep{Spezzi_2013} were confirmed as cluster members by \citet{Galli_2020b}.  We report these objects as "Unobserved" members in Table~\ref{tab:physical_param}. 

We note that all the Class~II sources not observed in our survey but confirmed as member by \citet{Galli_2020b} were observed but not detected at 70\,$\mu$m \citep{Spezzi_2013}. In Appendix~\ref{apdx:comparison}, we show that, out of the 10 disks undetected at 70\,$\mu$m included in our sample, only 3 were detected with our ALMA observations. It is thus likely that most of the unobserved Class~II sources would also not have been detected with our observations.

\subsection{ALMA observations}

Our ALMA observations (Project 2013.1.00708.S, PI:~M\'enard) were obtained on 2015 August 27, 
with an array configuration made of 40 antennas with baselines ranging from 26 to 1170 m. 
The continuum spectral windows were centered on~234.2\,GHz and 217.2\,GHz, 
giving a mean continuum frequency of~225.7\,GHz~(1.3\,mm). 
The other two spectral windows were set up to include three CO isotopologue lines. They covered the~$^{12}$CO,~$^{13}$CO, and C$^{18}$O $J=2-1$ transitions at~230.538\,GHz, 220.399\,GHz, and 219.56\,GHz. Each spectral window had~0.33~km\,s$^{-1}$ velocity resolution. 
The integration time was 2.5 min on-source per target giving an average continuum RMS of 0.17~mJy\,beam$^{-1}$. We used the calibration script provided by the observatory, with CASA~\citep{McMullin_2007} version 4.3.1, to calibrate the raw data.

\begin{figure*}[htbp]
\centering
\includegraphics[width =17.5cm]{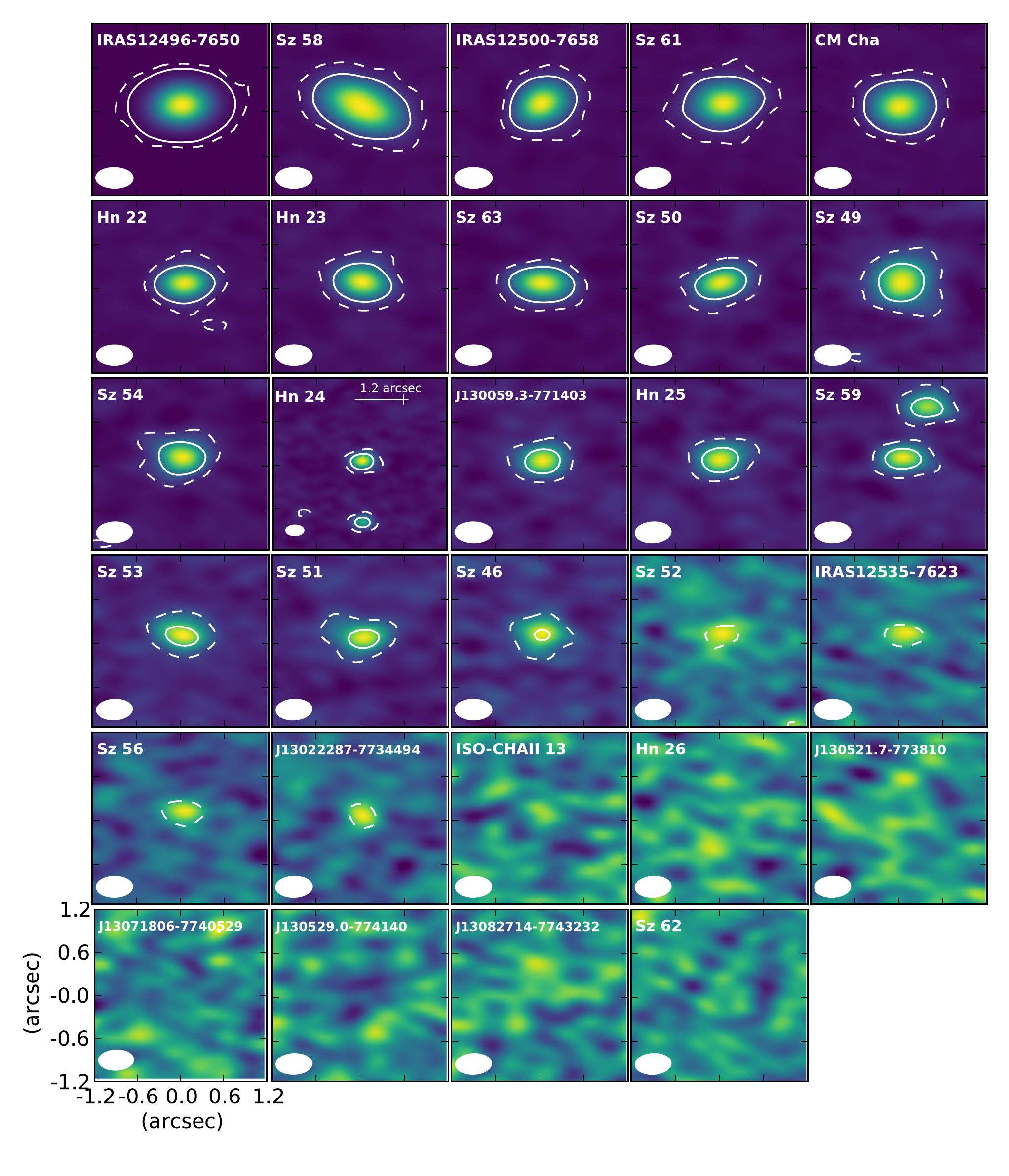}
\caption{Continuum images ordered by decreasing fluxes. Images are 2.4$\arcsec{}\times$2.4\arcsec{}, except for Hn 24 where we use 4.8\arcsec{}$\times$4.8\arcsec{} to show the secondary source. The beam is shown in the lower left corner of each panel. 3$\sigma$ and 15$\sigma$ contours are shown in solid and dashed lines, respectively. The color map is chosen to go from from -3$\sigma$ to the maximum intensity of each image.}
\label{fig:continuum-images}
	\vspace{3cm}
\end{figure*}

We produced the continuum images from the calibrated visibilities over the continuum channels by using the CASA \texttt{clean} function with a Briggs robust weighting parameter of +0.5. 
To maximize the dynamic range of the brightest sources, we performed a phase-only self-calibration on CM\,Cha, Hn\,22, Hn\,23, IRAS12500-7658, Sz\,58, and Sz\,61. For these sources, we used solution intervals of the scan length~("inf") and combined all spectral windows.
In the case of the brightest target, IRAS12496-7650 (also called DK\,Cha), phase and amplitude self-calibration were performed.  The first two iterations were phase only (solution intervals of~"inf" and~6.05s), and third round was an amplitude and phase calibration with a solution interval of the duration of the whole scan. 
We present the continuum images in Fig.~\ref{fig:continuum-images}. They achieve an averaged angular resolution of~$0.48\arcsec\times0.25\arcsec$ ($95\times50$\,au).

We extracted $^{12}$CO, $^{13}$CO, and  C$^{18}$O channel maps from the calibrated visibilities after subtracting the continuum from the spectral windows containing line emission using the \texttt{uvcontsub} routine in CASA. For the brightest sources, we also applied the continuum self-calibration solutions to the gas line data. We cleaned the sources with velocity channels of $0.35\ \text{km\,s$^{-1}$}$, and with a Briggs robust weighting parameter of~0. 
We obtain an average angular resolution of $0.51\arcsec\times0.28\arcsec$ ($100\times55$\,au) for the CO lines.

\begin{figure*} [htbp]
\centering
\includegraphics[width =17cm]{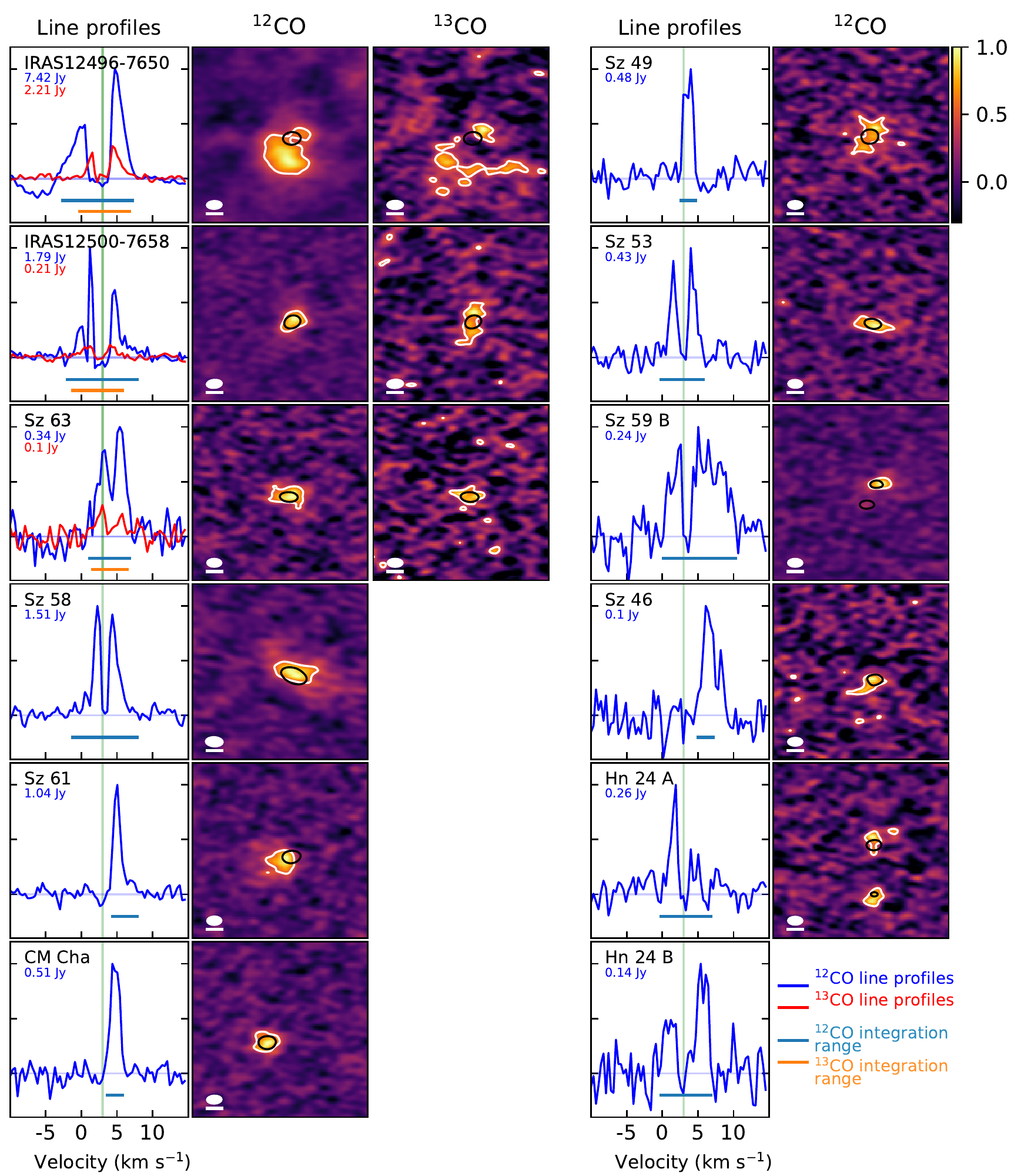}
	\caption{Line profiles (left panels), $^{12}$CO normalized moment 0 maps (middle panels), and $^{13}$CO normalized moment 0 maps (right panels) for the sources detected in  $^{12}$CO. Each line displays two sources. For each source, we display the continuum contours at 50\% of the continuum peak (black line) and the CO contours at 50\% of the CO peak (white line) on the CO moment 0 maps. The beam size is shown in the bottom left corner of each panel, along with a white line representing a 0.5\arcsec scale. On the line profiles plots (left panels), we also show a green vertical line at 3~km\,s$^{-1}$, where the cloud absorption is estimated, and the size of the integration range to estimate the $^{12}$CO and $^{13}$CO fluxes reported in Table~\ref{tab:table_CO}. The maximum fluxes of each line profile are displayed in the top left side of the panels, colored in blue for $^{12}$CO and in red for $^{13}$CO. The y ticks of the line profiles mark 0\%, 50\%, and 100\% of the maximum fluxes. }
	\label{fig:mom0}
	\vspace{3cm}
\end{figure*}

\begin{table*} 
\caption{$1.3$\,mm continuum properties. }
	\centering
    \begin{tabular}{lrcccrc}
\hline\hline
Source &$F_{1.3\mathrm{mm}}$ &   $a_\mathrm{1.3mm}$ & $i$& PA& $M_\mathrm{dust}$~~~~~\\ 
 &   (mJy) & (mas)& {(deg)} &  (deg) & (M$_\oplus$)~~~~~\\
\hline
J13022287-7734494 &    0.8 $\pm$ 0.2 & ... &... &...  &  1.7 $\pm$ 0.3 \\
J13071806-7740529 &    <0.5  &  ... & ... & ... &  <0.9 \\
J13082714-7743232 &     <0.5  &  ... &... &...& <1.1 \\
CM\,Cha & 38.8 $\pm$ 0.2  &  320 $\pm$ 5  &  77 $\pm$ 9 & 85 $\pm$ 22 &  34.1  $\pm$ 0.2 \\
Hn\,22&     21.8 $\pm$ 0.2 &   175 $\pm$ 6  &   51 $\pm$ 3  &   -52 $\pm$ 5  &  28.1  $\pm$ 0.2 \\
Hn\,23 &    20.7 $\pm$ 0.2  &   190 $\pm$ 4   &  $\times$ & -22 $\pm$ 57 & 17.7  $\pm$ 0.2 \\ 
Hn\,24\,A &     10.5 $\pm$ 0.2  &  193 $\pm$ 15  &    $\times$  & 67 $\pm$ 57 & 8.4  $\pm$ 0.2 \\
Hn\,24\,B & 4.8 $\pm$0.2  & ...&...&  ...&   5.5  $\pm$ 0.2 \\
Hn\,25 &      5.4 $\pm$ 0.2  & 195 $\pm$ 15&  $\times$ & -16 $\pm$ 8 &  6.2  $\pm$ 0.2 \\
Hn\,26 &     <0.5  & ... &  ...& ...& < 0.6 \\
IRAS12496-7650&    708.6 $\pm$ 0.3  &  354 $\pm$ 1  &   25 $\pm$ 1  &   125 $\pm$ 1  & 337.5  $\pm$ 0.1 \\
IRAS12500-7658&    49.6 $\pm$ 0.3 &  258 $\pm$ 5  &    $\times$  & -84 $\pm$ 57 &  266.5  $\pm$ 1.4 \\
IRAS12535-7623 &   1.0 $\pm$0.2  & ...&...& ...& 0.7  $\pm$ 0.1\\
ISO-CHAII 13 &   <0.5  &...  &... & ...& < 2.3 \\
J130059.3-771403 &      5.7 $\pm$ 0.2  & 188 $\pm$ 15 &  $\times$ & 4 $\pm$ 8 &    6.7  $\pm$ 0.2 \\
J130521.7-773810 &     <0.5  &... &  ...& ... & < 0.6 \\
J130529.0-774140 &    <0.5  &... & ...  & ...& < 0.6 \\
Sz\,46 &     2.8 $\pm$ 0.2  & ... &... & ... & 3.2  $\pm$ 0.2 \\
Sz\,49 &    10.9 $\pm$ 0.3  &    401 $\pm$ 14  &  39 $\pm$ 5  &   -1  $\pm$ 7  &   14.6  $\pm$ 0.3 \\ 
Sz\,50 &    11.0 $\pm$ 0.2  &   169 $\pm$ 15  &  $\times$ &   64 $\pm$ 57 &    5.7  $\pm$ 0.1 \\
Sz\,51  & 4.3 $\pm$ 0.2  &  249 $\pm$ 33  &  58 $\pm$ 8  &   -70  $\pm$ 10  &  4.5  $\pm$ 0.2 \\
Sz\,52 &   0.7 $\pm$ 0.2  & ... & ...&... & 1.0 $\pm$ 0.3 \\
Sz\,53  &     4.8 $\pm$ 0.2  &    307 $\pm$ 32  &  57 $\pm$ 6 &   65 $\pm$ 8  & 5.6  $\pm$ 0.2 \\
Sz\,54 &   10.5  $\pm$ 0.2  & 224 $\pm$ 9  &  54 $\pm$ 7  &  12 $\pm$ 6  & 7.0  $\pm$ 0.1 \\ 
Sz\,56 &  1.0  $\pm$ 0.2 & ... &...&... & 1.1  $\pm$ 0.2 \\
Sz\,58   &    60.7 $\pm$ 0.3  &  699 $\pm$ 5  &  60 $\pm$ 1  &    61 $\pm$ 1  &  53.9  $\pm$ 0.3 \\
Sz\,59\,A&  4.8 $\pm$ 0.2  & ... & ...& ...&  3.9  $\pm$ 0.1 \\
Sz\,59\,B &   4.1 $\pm$ 0.2 &...& ...& ...& 4.7  $\pm$ 0.2\\
Sz\,61  &    44.8 $\pm$ 0.2  &   445 $\pm$ 9  &   42 $\pm$ 2  & -73 $\pm$ 3 &  35.1  $\pm$ 0.2 \\
Sz\,62 &  <0.5  &  ...& ... &...& < 0.7 \\
Sz\,63 &    20.4  $\pm$ 0.2  & 343 $\pm$ 9  &  67 $\pm$ 1  &    83 $\pm$ 1  &   27.1  $\pm$ 0.3 \\
\hline
\end{tabular}
	\tablefoot{Gaussian (point source) models were fitted in the uv plane for resolved (unresolved) sources. We report the continuum fluxes or 3$\sigma$ upper limits~(F$_{1.3\mathrm{mm}}$), the resolved major axis FWHM~($a_{\mathrm{1.3mm}}$), inclination ($i$), the position angles (PA), and the estimated dust masses (Sect. \ref{sec:dust_mass}). The flux and dust mass uncertainties do not include the typical 10\% calibration uncertainty.  We also note that six sources were only resolved in one direction so their inclinations and PA could not be accurately evaluated. We indicate these sources by the symbol "$\times$" in the inclination column.  }
	\label{tab:cont-emission}
\end{table*}

Using the channel maps, we generated moment 0 maps for the detected sources. We used the \texttt{immoments} CASA task and generated the map with all the spectral channels where the source is visually detected. 
In addition, we used the CO channel maps to generate line profiles for each source and isotopologue, over a range from -10 to +15\,km\,s$^{-1}$. The line profiles of each source are integrated over a unique spatial range for all channels, the size of which depends on the detectability of the corresponding isotopologue. For the sources that are detected in at least one spectral channel, the integrating area corresponds to the ellipse that encompasses all pixels (in all channels) above a 3$\sigma$ limit. We define $\sigma$ as the global rms of the channel maps where the source is not detected. 
On the other hand, when the sources are not detected in any channel, we used a square of size 1\arcsec{}$\times$1\arcsec{} (close to the mean size of the detected sources), centered near the phase center to extract the spectrum. $^{12}$CO and $^{13}$CO line profiles, and moment~0 maps of the detected sources are shown in Fig.~\ref{fig:mom0}.

\begin{table}
\caption{Integrated fluxes for the CO isotopologues derived from the line profiles.}
\small
\centering
\begin{tabular}{lccc}
\hline \hline
Source & F$_{\,^{12}\mathrm{CO}}$  & F$_{\,^{13}\mathrm{CO}}$&  F$_{\,\mathrm{C}^{18}\mathrm{O}}$\\
 & (Jy\,km\,s$^{-1}$) & (Jy\,km\,s$^{-1}$) & (Jy\,km\,s$^{-1}$) \\
\hline
J13022287-7734494 & $< 0.14$ & $< 0.14$ & $< 0.12$\\
J13071806-7740529 & $< 0.15$  & $< 0.14$&$< 0.10$\\
J13082714-7743232 & $< 0.11$ & $< 0.16$& $< 0.10$\\
CM Cha  & $0.67 \pm 0.04$ & $< 0.19$& $< 0.09$ \\
Hn 22  & $< 0.13$ & $< 0.15$& $< 0.10$ \\
Hn 23  & $< 0.13$& $< 0.14$ &$< 0.10$ \\
Hn 24 A & $0.35 \pm 0.04$ & $< 0.15$&  $< 0.11$\\
Hn 24 B& $0.33 \pm 0.04$ & $< 0.15$ & $< 0.11$ \\
Hn 25 & $< 0.15$ & $< 0.14$&$< 0.11$\\
Hn 26 & $< 0.13$& $< 0.13$&$< 0.12$ \\
IRAS12496-7650  & $20.58 \pm 1.66$ & $5.26 \pm 0.33$ & $< 0.16$ \\
IRAS12500-7658  & $~~2.93 \pm 0.22$ & $0.60 \pm 0.06$&$< 0.14$ \\
IRAS12535-7623 & $< 0.15$ & $< 0.15$&$< 0.10$ \\
ISO-CHAII 13 & $< 0.14$&$< 0.15$ &$< 0.09$\\
J130059.3-771403 & $< 0.15$ & $< 0.16$ &$< 0.11$ \\
J130521.7-773810& $< 0.16$ & $< 0.14$ &$< 0.10$ \\
J130529.0-774140 &$< 0.13$& $< 0.14$ &$< 0.11$\\
Sz\,46 & $0.12 \pm 0.01$ & $< 0.13$ & $< 0.11$ \\
Sz\,49 & $0.61 \pm 0.04$ & $< 0.17$&$< 0.10$ \\
Sz\,50 & $< 0.17$& $< 0.17$ &$< 0.13$ \\
Sz\,51 & $< 0.14$  & $< 0.15$ &$< 0.09$ \\
Sz\,52 & $< 0.13$ & $< 0.16$&$< 0.12$ \\
Sz\,53 & $0.84 \pm 0.06$ & $< 0.15$ & $< 0.10$ \\
Sz\,54 & $< 0.14$ & $< 0.12$ & $< 0.11$\\
Sz\,56 &  $< 0.11$  &$< 0.13$ &$< 0.09$\\
Sz\,58  & $4.01 \pm 0.24$ & $< 0.15$ &$< 0.11$\\
Sz\,59 A & $<0.16$& $< 0.14$ &$< 0.11$ \\
Sz\,59 B& $1.14 \pm 0.09$ & $< 0.14$ &$< 0.11$ \\
Sz\,61  & $1.31 \pm 0.08$ & $< 0.16$&$< 0.11$ \\
Sz\,62 & $< 0.14$ & $< 0.14$& $< 0.11$ \\
Sz\,63 & $1.04 \pm 0.07$ & $0.21\pm 0.03$&$< 0.12$ \\
\hline
\end{tabular}
\tablefoot{The upper limits for the non-detections correspond to three times the RMS of the line profile, integrated over a line width of 6.3~km\,s$^{-1}$.}
\label{tab:table_CO}
\end{table}
 
\section{Results}
\label{sec:results}
\subsection{Continuum emission}
\label{sec:observation_cont}
We measured the continuum emission by fitting an elliptical Gaussian model to the visibility data, using the CASA \texttt{uvmodelfit} task. This model has six free parameters: integrated flux density~($F_\mathrm{\mathrm{1.3mm}}$), full width half maximum (hereafter FWHM) along the major axis~($a_\mathrm{1.3mm}$), aspect ratio of the axes~($r$), position angle~(PA), and the right ascension and declination of the phase center ($\Delta \alpha$, $\Delta\delta$). 
If the ratio of $a_\mathrm{1.3mm}$ to its uncertainty is less than five, we fitted the visibilities with a point source model with only three free parameters~($F_\mathrm{1.3mm}$, $\Delta \alpha$, $\Delta\delta$).
Table~\ref{tab:cont-emission} gives the measured 1.3\,mm continuum fluxes. For the sources fitted with an elliptical Gaussian, we also report $a_\mathrm{1.3mm}$ and PA, as well as the inclination, $i$, estimated from $r$ assuming that the disks are azimuthally symmetric. For the detected sources, the phase centers from the fitting are also reported in Table~\ref{tab:physical_param}.

Out of 31 sources, 24 are detected above a 3$\sigma$ significance threshold. 
This includes two secondary sources, which are detected in the fields of Sz\,59 and Hn~24. We measured a separation of 0.70\arcsec{} and PA of -25$^\circ$ for Sz\,59 from the continuum fits, which is consistent with the measurement by \citet{Geoffray_2001}. 
Hn~24~B is a new companion candidate, since it is not referenced in the literature. From the continuum fit, we measured a separation of 1.67\arcsec{} and a PA of~0$^\circ$.

Additionally, our results indicate that 16 sources are also resolved in at least one direction. For six of them, even if the major axis size is well resolved by the elliptical Gaussian model, the disk was not resolved in the other direction, which implies that the inclination and position angle could not be accurately evaluated. We indicate these disks by the symbol "$\times$" in the inclination column of Table~\ref{tab:cont-emission}. Observations at higher angular resolution are needed to estimate the inclination of these systems.
 
\subsection{CO line emission}
\label{sec:observation_CO}

We measured the line fluxes by integrating the line profile over the spectral range where the source is detected by more than 3$\sigma$. 
For the detected sources, the mean line width is $\sim$6.3~km\,s$^{-1}$. We represent the line width by horizontal lines at the bottom of each (left) panels of Fig.~\ref{fig:mom0}. For the detected sources, we estimated the flux error as the RMS of the line profile outside the source, integrated over the width of the emission. 
On the other hand, we report upper limits for the non-detections. They correspond to three times the RMS of the line profile, integrated over a line width of 6.3~km\,s$^{-1}$. 
We present the integrated fluxes and uncertainties for the three isotopologues in Table~\ref{tab:table_CO}.

Out of 31 targets observed, 12 are detected in $^{12}$CO, 3~in~$^{13}$CO, and none in C$^{18}$O. This includes the two secondary sources that are detected in $^{12}$CO only. 
All the sources detected in $^{13}$CO are also detected in $^{12}$CO, and all the sources detected in $^{12}$CO are detected in the continuum. Each source detected in a CO isotopologue is also spatially resolved in this isotopologue line. 
In the particular case of IRAS12496-7650, the $^{12}$CO and $^{13}$CO emissions appear to extend up to the maximum recoverable scale of the observations, which suggests that part of the emission is possibly filtered out. In addition,  we caution the reader for the presence of significant foreground absorption for all sources.
From the moment 0 maps displayed in Fig.~\ref{fig:mom0}, we see that the line emission of some sources is not centered on the continuum~(e.g., Sz\,61), which suggests that we are missing part of the emission for each of them. Furthermore, some line profiles are also asymmetric~(e.g., CM\,Cha) and/or have minima~(absorption) that go down to the continuum level~(e.g., IRAS12500-7658). This is not compatible with a Keplerian profile without absorption. The large cloud absorption appears to be located around $\sim$3\,km\,s$^{-1}$ (green line on Fig.~\ref{fig:mom0}), which is compatible with the study of~\citet{Mizuno_2001}. There is possibly some dispersion in the cloud velocity since some sources do not show significant absorption at the reported value (e.g., Sz\,63). 
The presence of significant cloud absorption indicates that the CO fluxes presented in Table~\ref{tab:table_CO} and the average line width of the profiles can be underestimated.

Additionally, from the black and white contours in the moment maps of Fig.~\ref{fig:mom0}, we find that the $^{12}$CO emission is systematically larger than its continuum counterpart. This result  has  been  observed  in various  other studies  \citep[e.g.,][]{van_der_plas_2017, ansdell_2018}, and can be explained by differences in optical  depth  between  gas  and  dust, or by the presence of grain growth and radial drift in the disks. However, due of the low angular resolution of our data (elongated beam) and significant foreground absorption, we could not get a reliable estimate the gas sizes. Higher angular resolution observations and detailed modeling of each object is needed to estimate quantitatively the dust and gas sizes, and to compare the dust and gas size ratio with previous studies~\citep[e.g.,][]{ansdell_2018, Facchini_2019, Trapman_2019, Sanchis_2021}. 

\begin{figure*}
	\centering
		\includegraphics[width=18.5cm]{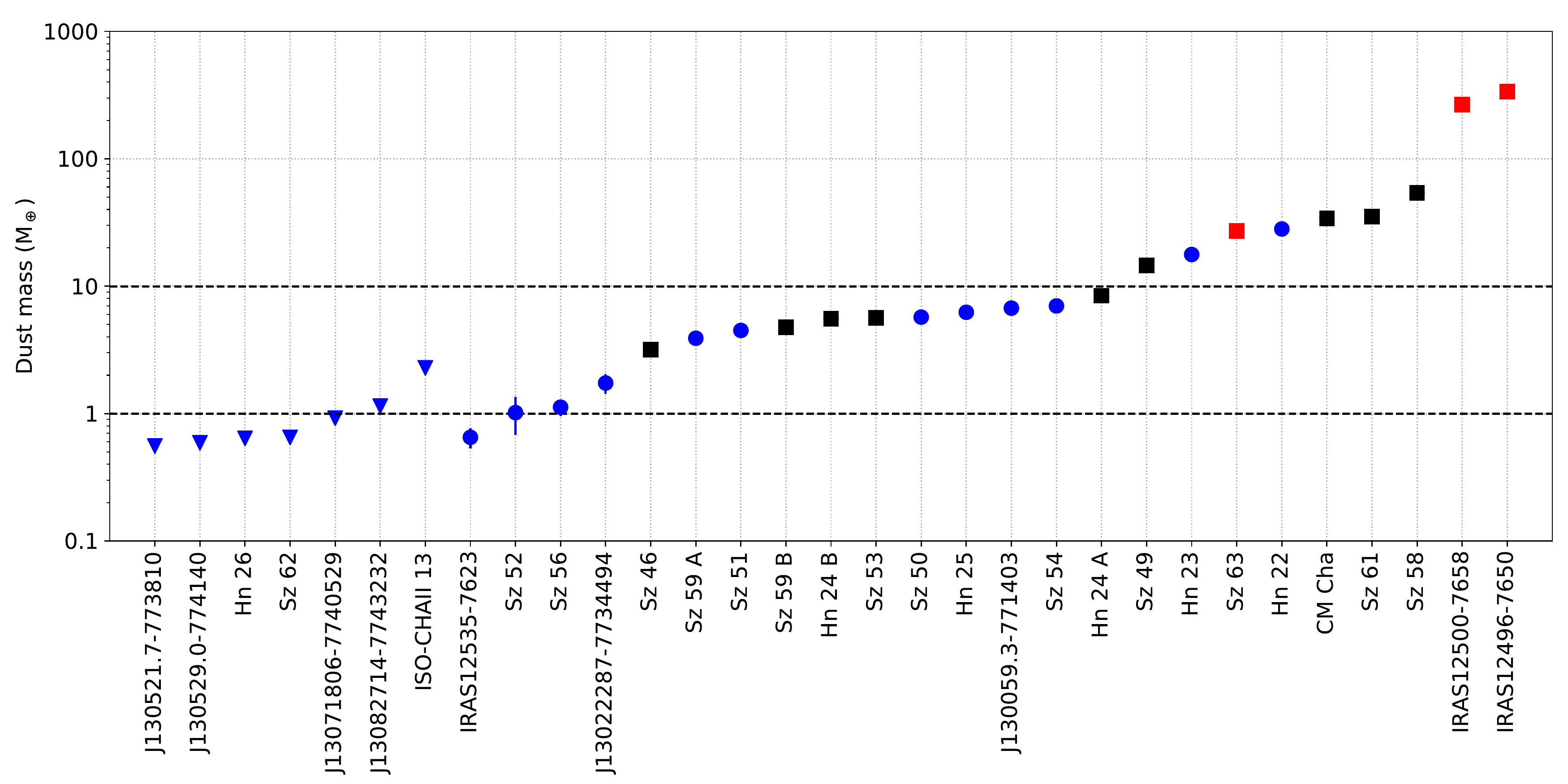}
		\caption{Dust masses for the 31 sources in our Cha~II sample expressed in Earth masses, ordered by increasing dust mass (Table~\ref{tab:cont-emission}). The black and red squares indicates the sources also detected in $^{12}$CO and $^{13}$CO, respectively. Round symbols show continuum only detected sources and the downward-facing triangles correspond to 3$\sigma$ upper limits for non-detections.}
		\label{fig:Mass_ChaII}
\end{figure*}

\section{Disk properties}
\label{sec:prop_dust}
\subsection{Dust masses}
\label{sec:dust_mass}

Assuming that the continuum emission is optically thin at~1.3\,mm, it is possible to infer the disk dust mass ($M_\mathrm{dust}$) from the continuum millimetric flux ($F_\nu$) at a given wavelength~\citep[e.g.,][]{Hildebrand_1983}:
\begin{equation}
M_\mathrm{dust} = \frac{F_\nu d^2}{\kappa_\nu B_\nu(T_\mathrm{dust})}.
\label{eq:equation_masse}
\end{equation}

We assumed a grain opacity $\kappa_\nu$ of $2.3\ \mathrm{cm}^2.\mathrm{g}^{-1}$ at 230~GHz~\citep{Beckwith_1990}, with $\kappa_\nu~\propto~\nu^{0.4}$~(as in other studies, e.g.,  \citealt{Andrews_2013, Pascucci_2016}, and consistently with recent integrated spectral index measurements, e.g., \citealt{Ribas_2017}). We used the individual \emph{Gaia} distances of each object as reported in Table~\ref{tab:physical_param}. 
When the sources do not have \emph{Gaia} parallaxes, we used the average distance of the well characterized objects:~198\,pc~\citep{Dzib_2018, Galli_2020b}. 
We also adopted the relationship of $T_\mathrm{dust}$ with $L_\star$ from~\citet{Andrews_2013}, inferred with a grid of radiative transfer models: $T_\mathrm{dust}= 25 \mathrm{K} \left( L_\star / L_\odot\right) ^{1/4}$. 
We used the luminosities determined by~\citet{Spezzi_2008}, rescaled to the \emph{Gaia} distances. For the sources that were not characterized spectrally (Hn 24 B, J130059.3-771403, J130521.7-773810, J130529.0-774140, and Sz 59 B), we applied a characteristic dust temperature of T$_\mathrm{dust}=20$\,K~\citep{Andrews_Williams_2005}.

Fig.~\ref{fig:Mass_ChaII} shows the detections and upper limits of our dust mass estimates. The values are reported in Table~\ref{tab:cont-emission}. They range from~$\sim$0.7\,M$_\oplus$~(IRAS12535-7623) to~$\sim$337.5\,M$_\oplus$ (IRAS12469-7650). 
We note, however, that the scaling relation between T$_\mathrm{dust}$ and L$_\star$ was calibrated for luminosities larger than~0.1\,L$_\odot$. In our sample, five objects have luminosities lower than this value, including IRAS12500-7658 and ISO-CHAII 13 that have luminosities close to  0.01~L$_\odot$. For these two sources, the dust mass is probably overestimated by a factor of $\sim$2~\citep{van_der_plas_2017}.

\subsection{Comparison with other regions}
\label{sec:Luminosity_function}

Over the last years, observations of nearby star-forming regions have shown a trend for disk dust mass to decrease with age, which can be a sign for disk dissipation and/or dust evolution~\citep[e.g.,][]{Ansdell_2016}. 
In this section, we expand the study of cumulative dust mass distribution to the Cha~II star-forming region, aiming to add constraints on the evolution of dust mass with time. We focus our analysis on isolated, low mass star-forming regions (see Sect.~\ref{sec:method_distrib}) for which the dissipation is likely not affected by external factors.

Previous studies have found that the dust mass (M$_\mathrm{dust}$) correlates with the stellar mass~\citep[M$_\star$, e.g.,][]{Andrews_2013, Pascucci_2016, Ansdell_2017, Cazzoletti_2019}. 
Because of this relation, low mass stars are expected to host less massive dust disks than more massive stars. This implies that the comparison of dust mass distributions of star-forming regions with different stellar mass distributions might lead to inadequate conclusions. 
In this context, some of our analysis consists in comparing the dust-to-stellar mass ratio distributions for different regions~\citep[see also][]{Barenfeld_2016}. 
Using this ratio allows us to reduce the impact of potentially different M$_\star$ distributions and, to first order, to study the evolution of M$_\mathrm{dust}$ as a function of time. 
In addition, for comparison with previous studies, we also present the dust mass distributions for the same regions.

\subsubsection{Sample}
\label{sec:method_distrib}

In this analysis, we consider seven star-forming regions observed at millimeter wavelengths and for which stellar masses can be well estimated. They are Upper\,Sco~\citep{Barenfeld_2016}, Cha~II~(this study), IC\,348~\citep{Ruiz-Rodriguez_2018}, CrA~\citep{Cazzoletti_2019}, Cha~I~\citep{Pascucci_2016},  Lupus~\citep[both band~7 and band~6 surveys;][]{Ansdell_2016, ansdell_2018}, and Taurus~\citep{Andrews_2013}.
We show the main characteristics of each region~(age, average distance, frequency of the observations) in Table~\ref{tab:regions}. Because objects of different SED classes are most likely in a different evolutionary stage, we selected only the Class II objects from all studies. For Upper\,Sco, they are the objects classified as "Full," "Transitional," and "Evolved" in Table~1 of \citet{Barenfeld_2016}. 

Additionally, we note that we did not include a number of other millimeter surveys in this analysis. This is either because they are located in dense or massive environments~\citep[e.g., $\sigma$-Orionis, ONC, OMC-2, NGC 2024, $\lambda$-Orionis;][]{Ansdell_2017, Eisner_2018, van_Terwisga_2019,van_terwisga_2020, Ansdell_2020} or because the stellar masses are not yet available~\citep[e.g., Ophiuchus, Lynds 1641;][]{Cieza_2018, Grant_2021}. We also did not include the SMA survey of the Serpens star-forming region~\citep{law_2017}, both because less than half of the known Class~II population of this region was observed and because the survey is significantly less sensitive compared to the other surveys of this study (lowest detected dust mass being $\sim$12\,M$_\oplus$).

\begin{table*}[h!]
\caption{Parameters of star-forming regions.}
\centering
\begin{tabular}{llcccccc}
\hline \hline
Region & Published age& Distance & Obs freq& M$_\mathrm{dust, min}$  &Detections/ & Median M$_\mathrm{dust}$&  Median M$_\mathrm{dust}/\text{M}_{\star}$\\
& (Myr)&  (pc)& (GHz)&(M$_\oplus$) &  Total sources &(M$_\oplus$)&($10^{-5}$)\\
\hline
     Upper\,Sco & 5 $-$ 11$^{(a, b)}$ & 145 $\pm$ 10$^{(b)}$ & 341.1 & 0.17 &53/75& 0.6 $\pm$ 0.2 & 1.9 $\pm$ 0.4\\
	  CrA & 1$-$3$^{(m)}$, 5$-$6$^{(n)}$ &  154 $\pm$ 4$^{(k)}$& 230.0& 0.08 &  16/26& 0.4 $\pm$ 0.1 & 1.4 $\pm$ 0.4\\
	 Cha II & 1 $-$ 2$^{(d)}$, 2 $-$ 6$^{(c)}$ & 198 $\pm$ 6$^{(k)}$ &225.7 & 0.65& 18/22& 4.5 $\pm$ 1.5 & 3.1 $\pm$ 1.5\\ 
     IC 348& 2 $-$ 3$^{(e)}$& 310 $\pm$ 20$^{(e)}$& 225.7& 1.52& 33/69& $<$ 1 & 5.4 $\pm$ 0.8\\
     Cha I & 1 $-$ 2$^{(d)}$, 2 $-$ 4$^{(f,g)}$ & 192 $\pm$ 6$^{(k)}$& 338.0 & 0.54 & 64/86&  3.5 $\pm$ 0.8 & 5.9 $\pm$ 1.4\\
	 Lupus b7 & 1 $-$ 3$^{(h, i)}$ & 160 $\pm$ 4$^{(k)}$& 335.8& 0.16&54/60& 6.1 $\pm$ 2.5 & 5.6 $\pm$ 0.7\\
	 Lupus b6& &&225.5 &0.59&60/68& 5.8 $\pm$ 1.5 & 5.1 $\pm$ 0.6\\
	 Taurus & 1 $-$2$^{(j)}$& 141 $\pm$ 7$^{(l)}$& 225.0 & 1.08 & 101/176& 4.4 $\pm$ 0.7 & 12.6 $\pm$ 2.8\\
	 \hline
\end{tabular}
\tablefoot{The column M$_\mathrm{dust, min}$ corresponds to the lowest dust mass detected in each star-forming region. The median M$_\mathrm{dust}$ and M$_\mathrm{dust}/\text{M}_{\star}$ correspond to the intersection of the cumulative distribution curves with the 0.5 value in the y-axis of the figures in Fig.~\ref{fig:distribution_functions}. This includes  both the detected disks and upper limits. }
\tablebib{$^{(a)}$~\citet{Pecaut_2012}, $^{(b)}$~\citet{herczeg_2015}, $^{(c)}$~\citet{Spezzi_2008}, $^{(d)}$~\citet{Galli_2020b}, $^{(e)}$~\citet{Ruiz-Rodriguez_2018}, $^{(f)}$~\citet{Pascucci_2016}, $^{(g)}$~\citet{luhman_2007},  $^{(h)}$~\citet{Ansdell_2016}, $^{(i)}$~\citet{Galli_2020c}, $^{(j)}$~\citet{Andrews_2013}, $^{(k)}$~\citet{Dzib_2018}, $^{(l)}$ \citet{Zucker_2019}, $^{(m)}$~\citet{Sicila-Aguilar_2011}, $^{(n)}$~\citet{Galli_2020}. }
\label{tab:regions}
\end{table*}

\subsubsection{Methods}
\label{sec:methods}

In order to provide a meaningful comparison of all regions, we recalculated both the dust and stellar masses in a consistent and homogeneous manner. 

\paragraph{Individual distances.}
We considered individual stellar distances. For Cha~II, we used the distances reported in Table~\ref{tab:physical_param}. We also excluded the sources classified as uncertain (including the two binary candidates), foreground, and background (Table~\ref{tab:physical_param}) from this analysis. For the other star-forming regions, whenever uncertainties are smaller than 10\%, we assigned the distance of the source from the \emph{Gaia~DR2} catalog~\citep{gaia_DR2_2018}. On the other hand, for sources with larger errors or that are not in the catalog, we used the average distance of the association~(see Table~\ref{tab:regions}).

\paragraph{Stellar masses.} We determined stellar masses for all data sets in a consistent way, using isochrones from~\citet{baraffe_2015} in the range 0.5 to 50 Myr. The tracks were interpolated to probe the mass range from 0.05 to 1.4~M$_\odot$, by steps of 0.01~M$_\odot$. We adopted the method described in~\citet{Andrews_2013} and~\citet{Pascucci_2016} to assign a stellar mass and an age. We first evaluated a likelihood function~\citep[Eq. 1 in][]{Andrews_2013} on each grid model,  assuming that the uncertainties in $\log (\mathrm{L}_\star / \mathrm{L}_\odot )$ and $\log (\mathrm{T}_\star/K)$ are 0.1 and 0.02, respectively, which correspond to the upper values for uncertainties in~\citet{Spezzi_2008}.
We then marginalized the distribution to estimate the stellar masses and their uncertainties, corresponding to the median, 18\%, and 84\% percentiles, respectively (see Table~\ref{tab:physical_param} for Cha~II).

We used stellar luminosities and temperatures from~\citet{Andrews_2013}, \citet{alcala_2017}, \citet{manara_2017}, \citet{luhman_2007}, \citet{Cazzoletti_2019}, \citet{Ruiz-Rodriguez_2018}, and~\citet{Barenfeld_2016} for Taurus, Lupus, Cha~I, CrA, IC\,348, and Upper\,Sco, respectively. Before estimating the stellar masses, we rescaled the luminosities to each individual stellar distance.

\paragraph{Dust masses.} We also recalculated dust masses in a homogeneous way from millimeter or submillimetrer fluxes \citep[from][]{Barenfeld_2016, Cazzoletti_2019, Ruiz-Rodriguez_2018, Pascucci_2016, Ansdell_2016, ansdell_2018, Andrews_2013} using Eq.~(\ref{eq:equation_masse}). We used the same temperature-luminosity relation and grain opacity as previously, and adopted the most recent distances.  
We report the mass of the least massive disk detected in each star-forming region in Table~\ref{tab:regions} (column M$_\mathrm{dust, min}$). 

We note that using the simplifying assumption of T$_\mathrm{dust}=20$\,K does not change the statistical significance of the results presented in this section. 
Also, it should be noticed that our analysis includes surveys observed at different frequencies, either ALMA band 6 ($\sim$225\,GHz) or band~7~($\sim$340\,GHz; see Table~\ref{tab:regions}). Using $\kappa_\nu \propto \nu$~\citep[as in other studies, e.g.,][]{Ansdell_2016, Cazzoletti_2019} instead of $\kappa_\nu \propto \nu^{0.4}$ \citep[this study and others, e.g.,][]{Andrews_2013, Pascucci_2016} corresponds to a difference of less than 25\% in the band~7 opacities. We checked that using either opacity law does not significantly affect the results. 

\paragraph{Disks sub-selection.} Because most of the samples used are only complete down to the brown dwarf limit, we considered only stars with derived masses above 0.1 M$_\odot$.   
Moreover, as \citet{baraffe_2015} tracks stop at M$_\star=1.4$\,M$_\odot$, we decided not to include stars where the fit of isochrone produces this value. We also omitted sources for which the stellar mass is not available, even if they have a measured dust mass. 
We present the number of sources considered in this study along with the number of detected sources in Table~\ref{tab:regions}. 

\begin{figure}
	\centering
        \includegraphics[width=9cm]{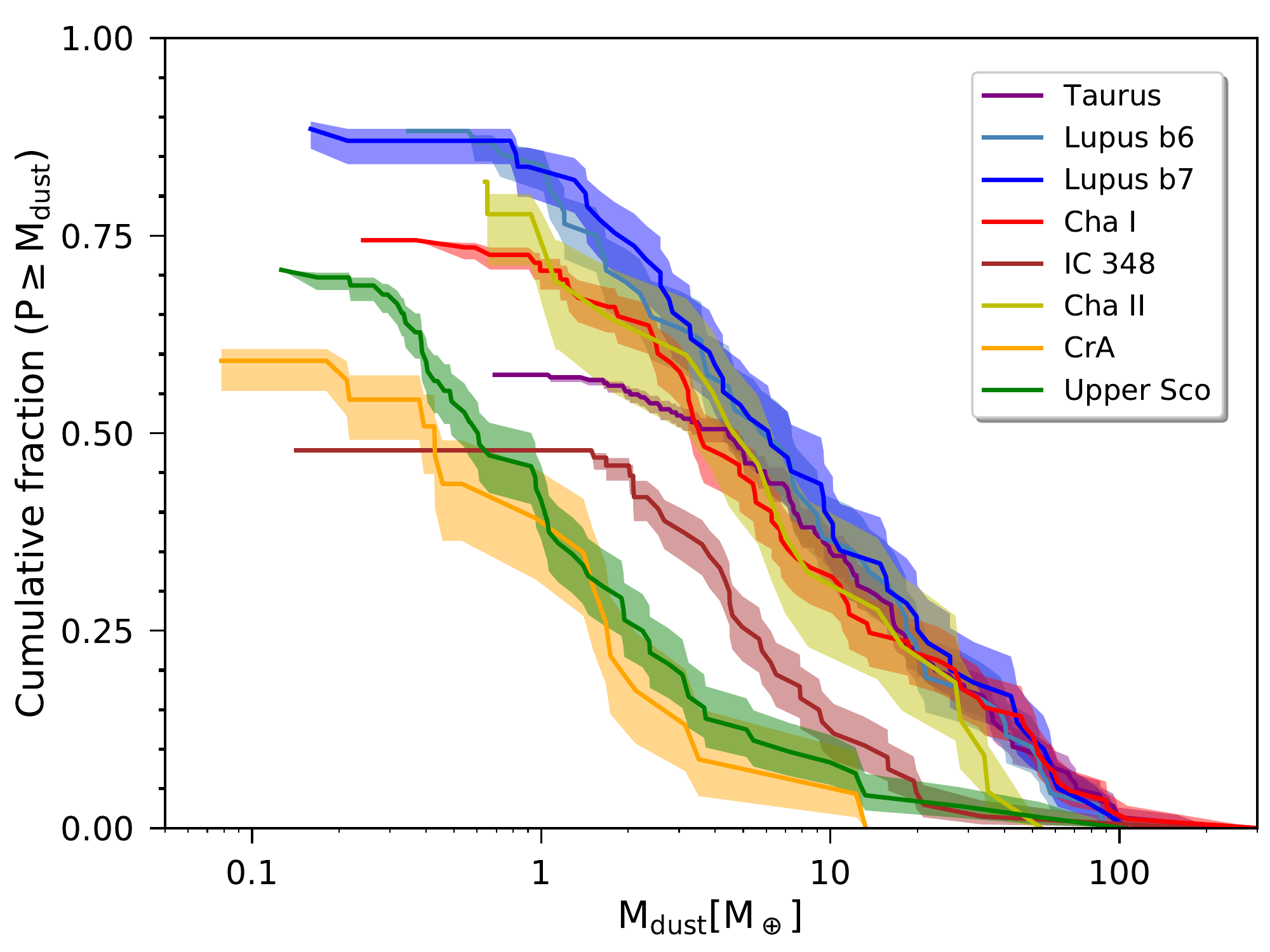}
        \includegraphics[width=9cm]{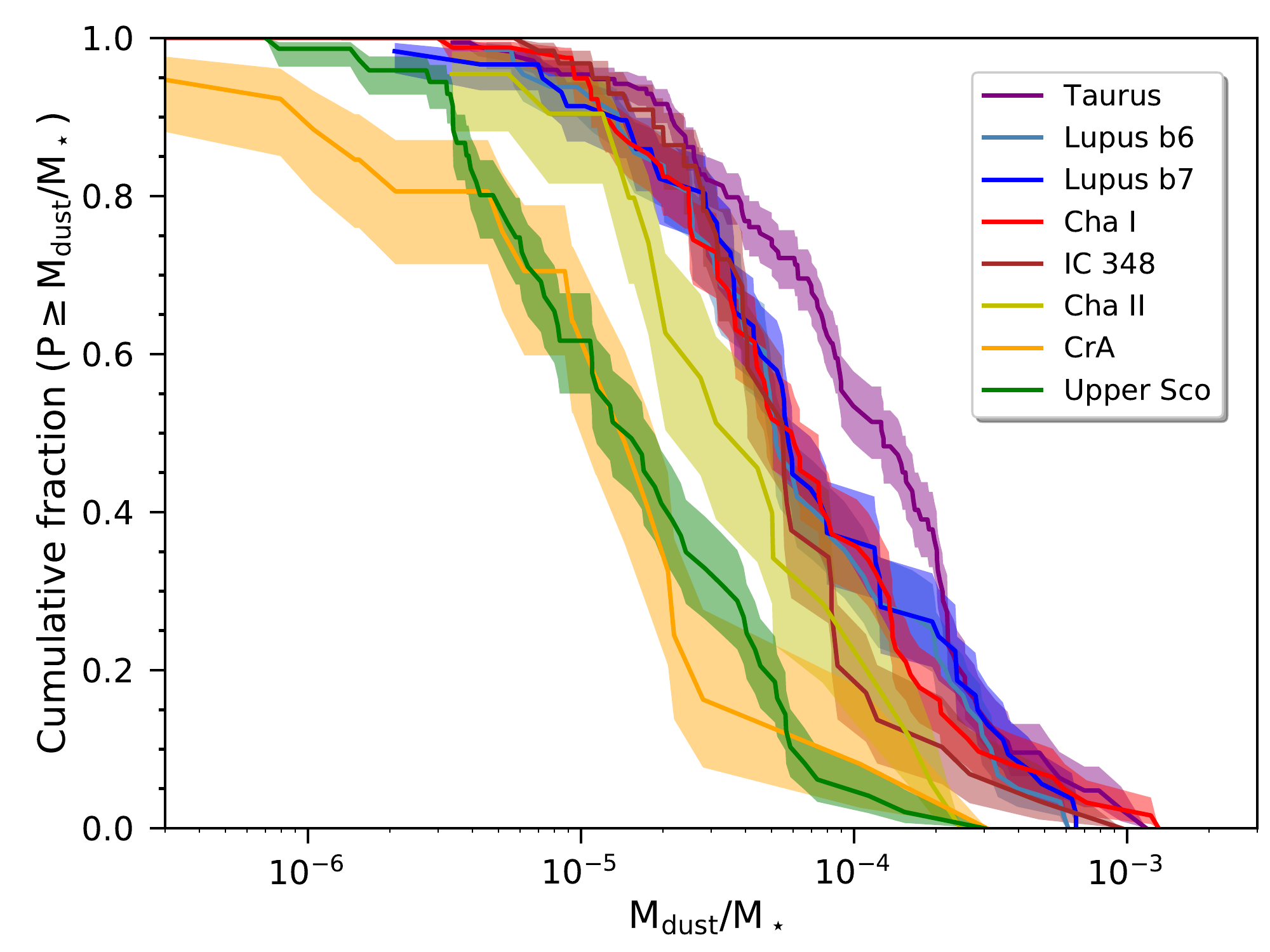}
 		\caption{Cumulative distributions functions generated by the Kaplan-Meier estimator. \emph{Top:} Cumulative distribution of the dust mass, normalized by the fraction of detected sources in each star-forming regions. \emph{Bottom:} Cumulative distribution of the dust-to-stellar mass ratio. The bottom distributions include disks not detected in millimeter emission. Those are scattered over the distribution of M$_\mathrm{dust}$/M$_\star$ (see Sect.~\ref{sec:cum_distrib}). We indicate 1$\sigma$ confidence intervals.}
		\label{fig:distribution_functions}
\end{figure}  

\subsubsection{Cumulative distributions}
\label{sec:cum_distrib}
In order to compare all star-forming regions, we generated two families of cumulative distributions. We used the Kaplan-Meier estimator in the {\texttt lifelines} package in Python\footnote{See documentation at http://lifelines.readthedocs.io/}, which takes into account upper limits and was used in previous studies~\citep[e.g.,][]{Ansdell_2016,law_2017}. We note that the Kaplan-Meier estimator assumes 
that the value of a censored point is precisely known~\citep[e.g.,][]{Feigelson_1985}. While the errors on M$_\mathrm{dust}$ are typically of a few percent (see Table~\ref{tab:cont-emission}), the uncertainties on the ratio of M$_\mathrm{dust}$/M$_\star$ are significantly larger (due to the larger uncertainty on M$_\star$; see Table~\ref{tab:physical_param}) and have large variations between sources, which is not taken into account with this estimator.

\paragraph{Dust masses.} First, as in previous studies, we estimated the cumulative distributions of M$_\mathrm{dust}$. They are shown in the top panel of~Fig.~\ref{fig:distribution_functions}, and scaled so that the maximum of each distribution corresponds to the fraction of detected sources in each sample.  
We find that most distributions, including that of Cha~II, have similar medians and shapes, but Upper\,Sco, CrA, and IC\,348 show a noticeable difference compared with the others, with median dust masses up to one order of magnitude smaller than in other regions for Upper\,Sco and CrA (see penultimate column of Table~\ref{tab:regions}). We note that when performing a parametric estimate of the dust mass distribution~\citep[e.g., in][]{Williams_2019}, we find similar results with the distributions of Upper Sco and CrA shifted to lower mass compared to the other regions. In addition to intrinsic differences in the dust mass, this might be related to different effects such as differences in stellar populations with other regions~(e.g., no stars between 0.9 and 1.4~M$_\odot$ in IC\,348) or to the scaling factor used for the M$_\mathrm{dust}$ distribution. The later might have to be modified if further studies identify new sources in the star-forming regions studied \citep[e.g.,][in CrA]{Galli_2020}, and if their observations at millimeter wavelength lead to a different fraction of detected sources in the samples.

\begin{table*}
\caption{Results of pair comparisons with the logrank test applied on the ratio M$_\mathrm{dust}$/M$_\star$. }
\centering
\begin{tabular}{ccccccccc}
\hline \hline
 & Upper Sco &  CrA & Cha II & IC348 & Cha I & Lupus B7 & Lupus B6 & Taurus\\
\hline
Upper Sco&&- & *&**&**&**&** & **\\
CrA&0.9&& -& ** & **& **& ** & **\\
Cha II&0.02&0.1&&-& -& *& - & **\\
IC348&7e-7&4e-4&0.3&&-& -&- & *\\
Cha I&< 1e-7&6e-05&0.1&0.5&&-&-&*\\
Lupus B7&< 1e-7&1e-4&0.045&0.5&1&&- & -\\
Lupus B6&< 1e-7&2e-4&0.06&0.8&0.8&0.4&&*\\
Taurus&< 1e-7&< 1e-7&2e-4&0.008&0.026&0.06&0.02&\\
\hline
\end{tabular}
\tablefoot{The lower left half of the table represents p-values for pairs comparisons, and the upper right half corresponds to the interpretation of the test. A p-value lower than 0.001 (**) corresponds to regions statistically distinct, p-values higher than 0.05 (-) to regions statistically similar, and p-values between 0.001 and 0.05 (*) to regions that are marginally different.}
\label{tab:logranktest}
\end{table*}

\paragraph{Dust-to-stellar mass ratios.} In order to limit the effect of different stellar populations, we also estimated cumulative distributions of dust-to-stellar mass ratio, which are shown in the bottom panel of~Fig.~\ref{fig:distribution_functions}. 
These curves go up to 1 rather than to the fraction of detected sources in each region even though disks with an upper limit on their dust mass are included. 
This is because of the known correlation of M$_\mathrm{dust}$ with M$_\star$. Indeed, while most non-detected systems have a lower dust mass than the detected disks (confining them to the lower mass end of the cumulative distribution of M$_\mathrm{dust}$ and justifying the scaling applied), most of these systems are found around low mass stars. 
We  checked that, on average, 60\% of the non-detected disks would have a M$_\mathrm{dust}/\mathrm{M}_\star$ ratio larger than three times the lowest ratio of a detected disk in the corresponding region when using the proper upper limits (only 6\% on average for M$_\mathrm{dust}$). So the non-detections would be scattered all over the cumulative distribution of M$_\mathrm{dust}/\mathrm{M}_\star$ and not only concentrated to the low end of the distribution, as it is the case for  M$_\mathrm{dust}$. If we were to perform more sensitive observations, these targets might be detected and have a large M$_\mathrm{dust}/\mathrm{M}_\star$, so the distribution of this ratio should not be normalized by the fraction of detected sources as opposed to the distribution of M$_\mathrm{dust}$.
We note again that the Kaplan-Meier estimator takes non-detections into account.

As highlighted by the relative position of each distribution, we find that most regions appear to have similar shapes, with the exception of Taurus. The Taurus distribution has a different shape than in the plot of M$_\mathrm{dust}$ because a large fraction of the non-detections are around low mass stars. This leads to a large number of entries with a large~M$_\mathrm{dust}$/M$_\star$ ratio, illustrating the effect previously mentioned.  
It is also clear the Upper Sco and CrA distributions have a shallower slope than the other distributions, as previously reported by~\citet{Ansdell_2017} and \citet{Cazzoletti_2019}. 
In the following subsection, we aim to statistically compare the different regions.

\subsubsection{Statistical test}
To test the statistical significance of the observed shift in the dust mass and mass ratio distributions, we performed two statistical tests on all star-forming regions. 
We used the logrank test, a non parametric method that compares the survival distributions of two samples, taking into account non-detections. The null hypothesis is that distributions of all regions are equal at all mass ratios. 

For the first test, we compared directly the distributions of M$_\mathrm{dust}$/M$_\star$ ratio and present the results in Table~\ref{tab:logranktest}. We find that Upper Sco and CrA are statistically different from all other regions except from Cha~II (marginal difference between Cha~II and Upper Sco), and they are statistically similar among themselves. Cha~II is also statistically different from Taurus, and the other regions are statistically similar or marginally different.

\begin{figure}
	\centering
        \includegraphics[width=9cm]{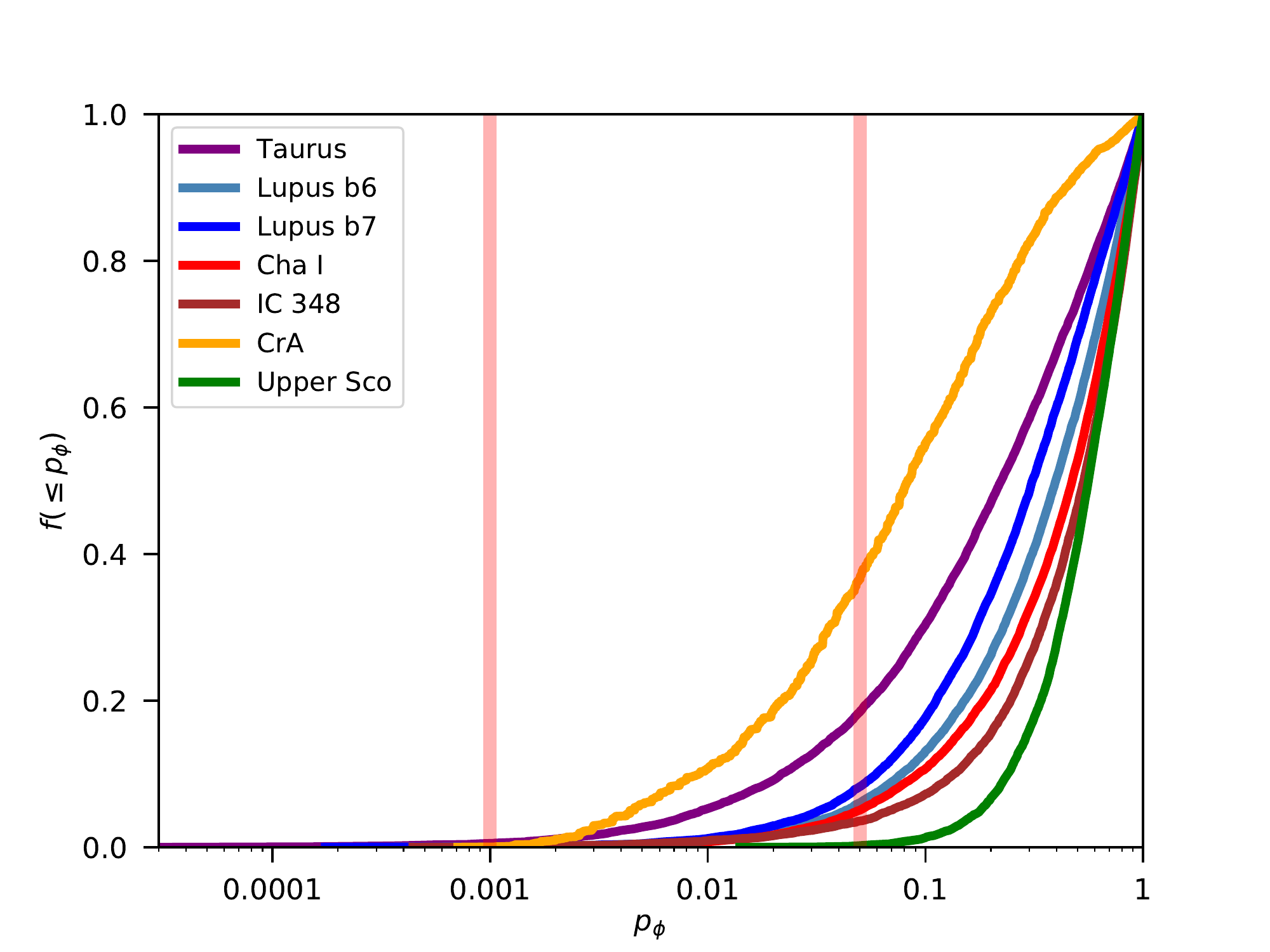}
 		\caption{Comparison of the dust mass distributions of different star-forming regions to that of Cha II. $p_\phi$ is the probability that the synthetic population drawn from the comparison samples and the reference sample come from the same parent population. $f(\leq p_\phi)$ is the cumulative distribution for $p_\phi$ resulting from the logrank two-sample test for censored data sets after $10^4$ MC iterations. The vertical lines correspond to $p_\phi$ of 0.05 and 0.001, respectively.}
		\label{fig:mc}
\end{figure}

We also performed a more robust statistical comparison of the dust mass distributions (M$_\mathrm{dust}$), following the methodology of \citet{Andrews_2013}. We first divided the stellar distributions into 5 mass bins between 0.1 and 1.4 M$_\sun$ and drew the same number of sources in each bin from the Cha II region (reference sample) and from other star-forming regions (comparison samples). Then, we performed a logrank test to test that the distributions are drawn from the same parent population. This process is repeated $10^4$ times for each compared region. We present the cumulative histogram of the results in Fig.~\ref{fig:mc}. As for Table~\ref{tab:logranktest}, a low $p_\phi$ value indicates that the regions are statistically different. We find median $p_\phi$ values of 0.22, 0.40, 0.31, 0.47, 0.54, 0.08, and 0.56, respectively for Taurus, Lupus (b6), Lupus (b7), Cha~I, IC\,348, CrA, and Upper Sco when compared to the Cha~II dust mass distribution. In other words, we find some consistency in the results with the first statistical test presented in this section: Cha~II appears to be statistically similar to Lupus, Cha~I, and IC\,348. However, in contrast to previous results from our first test performed on M$_\mathrm{dust}$/M$_\star$, we now find that the dust mass distribution (M$_\mathrm{dust}$) of the Cha~II star-forming region is also statistically similar to that of Upper Sco and Taurus, and is also potentially marginally different from CrA. The differences between the two statistical tests might indicate that the relationship of M$_\mathrm{dust}$ with M$_\star$ varies with the star-forming region considered, as found by previous studies~\citep[e.g.,][]{Ansdell_2017, Cazzoletti_2019}. Alternatively, as previously mentioned, we note that the Kaplan-Meier and logrank tests do not take into account the uncertainties on the censored values. Those can be large in the case of M$_\mathrm{dust}$ with M$_\star$ and may lead to an overestimation of the significance of the logrank test on M$_\mathrm{dust}$/M$_\star$.  We also note that the comparison of Cha~II with CrA was performed on a small number of disks ($<24$ in both regions) and might not be statistically significant.

If we consider that Cha~II is $4\pm2$\,Myr as found by \citet{Spezzi_2008}, we can interpret the shift in M$_\mathrm{dust}$ and M$_\mathrm{dust}$/M$_\star$ between the young regions (Taurus, Lupus, Cha I, and IC348), the intermediate Cha~II, and the older regions (Upper Sco and CrA) as an evolutionary effect: the older regions being less massive (see Table~\ref{tab:regions} and Fig.~\ref{fig:distribution_functions}) due either to a decline of dust mass with time or to dust evolution. When the age difference between two regions is large (e.g., $\geq$ 3 Myr), their distributions of M$_\mathrm{dust}$/M$_\star$ are statistically different and the distributions of M$_\mathrm{dust}$ in similar stellar mass bins are marginally different. On the other hand, regions of similar age are statistically similar. 
Alternatively, a recent study by \citet{Galli_2020b} suggested that the Cha~II region is younger, with an age around $1-2$ Myr. In that case, the shift in M$_\mathrm{dust}$ and M$_\mathrm{dust}$/M$_\star$ would not be related to disk evolution but possibly to different initial conditions between the different regions~\citep[see discussions in e.g.,][]{Cazzoletti_2019, Williams_2019}. However, the minor differences in the distributions of M$_\mathrm{dust}$ in similar stellar mass bins between the different regions (see Fig. ~\ref{fig:mc}) prevent us from drawing strong conclusions.

\subsubsection{Possible limitations}
\label{sec:limitation}

To convert the observed fluxes into dust masses, we assumed that disks were optically thin at 0.9\,mm and 1.3\,mm, meaning that the observed continuum flux is a reliable tracer of dust mass. This assumption may be partially incorrect since substructures are found to be ubiquitous in protoplanetary disks~\citep{Andrews_2018, Long_2018}, and often coincide with optically thick regions~\citep[e.g.,][]{Dullemond_2018, Dent_2019}. 
Several studies, for example comparing the dust masses estimated from radiative transfer or physical models with those calculated with Eq.~(\ref{eq:equation_masse})~\citep[e.g.,][]{Ballering_2019, Ribas_2020}, provide another indication that disks might be optically thick at 1.3\,mm. Indeed, they find that the analytical masses are generally underestimated~(by a factor of one to five) compared to the detailed results. 
Nevertheless, to facilitate the comparison with previous studies and because performing individual modeling of a large number of disks is extremely expensive, we assumed that the continuum flux is a reliable tracer of dust mass. 
Further surveys at longer wavelengths~\citep[e.g.,][]{Tazzari_2020}, expected to probe larger grains with lower millimeter opacities, will be useful to characterize with more details the decrease in dust mass with the age of the star-forming region.

Resolved and unresolved binary systems were also not filtered out from the samples. However, binary systems have been shown to disperse their disk faster than single star systems~\citep{cieza_2009, cox_2017, Zurlo_2020}, especially when their separation is smaller than 40~au \citep{kraus_2012}. 
We verified that only a very limited fraction of the disks included in this study are known close binaries~(r~<~40~au), less than 10\% in each region. 
Therefore, the multiplicity is likely to affect the statistics in all regions in similar way, and the  multiplicity is unlikely to affect the results unless it is a strong function of age.

Comparison between inhomogenous samples requires care. Here, all samples were observed with similar but yet different sensitivities. The cumulative distributions were generated using the Kaplan-Meier estimator that considers the upper limits of non-detections. In Appendix~\ref{sec:cumdistrib_bis}, we performed the analysis considering that each distribution had a similar disk mass detection limit. Although less statistically significant, the results are comparable with those presented above in this section. 
Thus, higher sensitivity observations may change only the very low mass end of the mass distribution.

Finally, the interpretation of the distribution functions from an evolutionary perspective presented here relies on the age of each individual region based on previous studies, in most cases prior to \emph{Gaia}. 
Although we used the new distances to reevaluate the stellar luminosities (and therefore masses) for each star, we did not reassess the age of each association.

   \section{Conclusions}
   \label{summary}
   
   We presented the first ALMA millimeter survey of 29 protoplanetary disks of the Chamaeleon~II star-forming region. We also detected two secondary sources in the fields of Hn 24 and Sz 59. 
   Our ALMA observations cover the 1.3 mm continuum as well as the $^{12}$CO, $^{13}$CO, and C$^{18}$O $J=2-1$ lines. 

   Out of our initial sample of 29 sources, we detect 22 disks in the continuum, 10 in $^{12}$CO, 3 in $^{13}$CO, and none in C$^{18}$O. We also detect the two companion candidates in the continuum and in $^{12}$CO. 
   We find that the $^{12}$CO emission is systematically larger than its continuum counterpart, which can be due to optical depth effects as well as radial drift and grain growth. 
   
   We also estimated the disk dust masses using the \emph{Gaia DR2} individual distances and find that the measured dust masses range from 337.5 M$_\oplus$ down to  0.7 M$_\oplus$. When accounting for the non-detections, we derived a median disk dust mass of 4.5 $\pm$ 1.5\,M$_\odot$ using a survival analysis. 
   We compared the dust mass distributions of our Cha~II sources with those of other isolated and low mass star-forming regions for which the stellar masses could be well estimated: Upper Sco, CrA, IC~348, Cha~I, Lupus, and Taurus. To limit the impact of potentially different distributions in stellar mass, we also compared the cumulative distributions of the dust-to-stellar mass ratio between all regions. 
   We find that the oldest region of the sample Upper Sco is statistically less massive than all other regions. Cha~II, whose age was recently revised from $4\pm2$ Myr~\citep{Spezzi_2008} to $1-2$ Myr by \citet{Galli_2020b} using the \emph{Gaia DR2} data release, is also statistically different from Taurus (Cha~II being less massive).
   All other regions are statistically similar when comparing their distributions of dust-to-stellar mass ratio. We also performed a second test, where we compare the dust mass distributions in similar mass bins for different regions. Similarly to the results of the first test, we find that Cha~II is statistically similar to Lupus, Cha~I, IC~348, but in contrast Cha~II is found to be statistically similar to Upper Sco and Taurus, and marginally different from CrA. 
 When considering the age of Cha~II as $4\pm2$ Myr, our results are consistent with a decline of the dust-to-stellar mass ratio with the age of the region or with dust evolution. On the other hand, if an age of $1-2$\,Myr is assumed, the shift in dust mass  might indicate differences in the initial conditions between regions. However, the minor statistical differences in dust mass as estimated on similar mass bins prevents us from drawing strong conclusions.
  Further surveys of intermediate age regions are crucial to understand the decrease of the dust mass with time. 
   
\begin{acknowledgements}
The authors thank the referee for their constructive report and useful suggestions that have significantly improved the paper. We also thank P. Galli for sharing sharing the membership tables for the Chamaeleon sources and S. van Terwisga for interesting discussions. This paper makes use of the following ALMA data: ADS/JAO.ALMA\#2013.1.00708.S. ALMA is a partnership of ESO (representing its member states), NSF (USA), and NINS (Japan), together with NRC (Canada), NSC and ASIAA (Taiwan), and KASI (Republic of Korea), in cooperation with the Republic of Chile. The Joint ALMA Observatory is operated by ESO, AUI/NRAO, and NAOJ. The National Radio Astronomy Observatory is a facility of the National Science Foundation operated under cooperative agreement by Associated Universities, Inc. MV, FM, MB, and GvdP acknowledge funding from ANR of France under contract number ANR-16-CE31-0013 (Planet Forming Disks). MV research was supported by an appointment to the NASA Postdoctoral Program at the NASA Jet Propulsion Laboratory, administered by Universities Space Research Association under contract with NASA. CC acknowledges support from DGI-UNAB project DI-11-19/R and ANID -- Millennium Science Initiative Program -- NCN19\_171. LC acknowledges support from ANID FONDECYT grant 1211656.  CP acknowledges funding from the Australian Research Council via FT170100040 and DP180104235. J.P.W. acknowledges support from NSF grant AST-1907486.
\end{acknowledgements}

\bibliographystyle{aa}
\bibliography{biblio}

 \begin{appendix}
 
 \section{Sample selection}
 \label{apdx:comparison}
    
\begin{figure*}
	\centering
        \includegraphics[width=18cm]{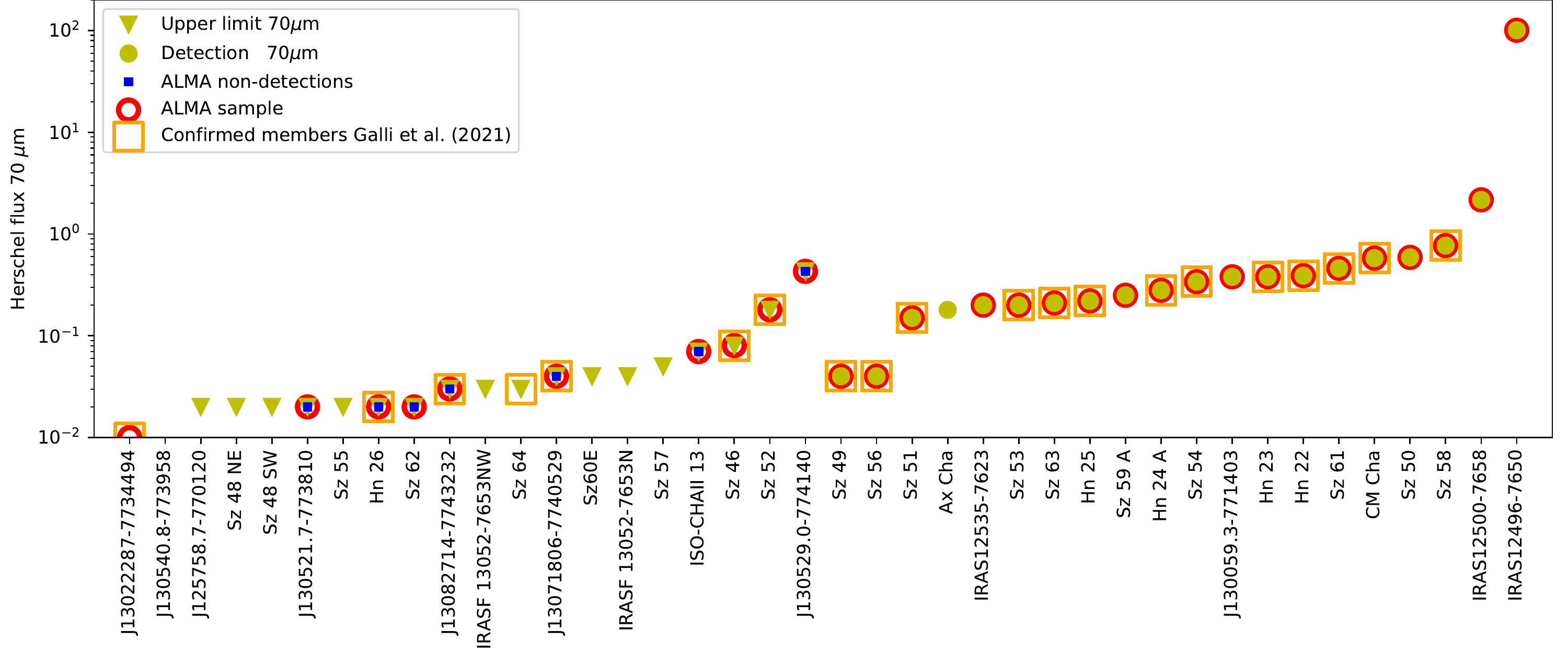}
 		\caption{{70\,$\mu$m flux from \citet{Spezzi_2013}. Circles indicate detected sources, and the triangles are the upper limits for the non-detections. Red circles indicate sources observed with our ALMA program, and blue squares show the sources that were not detected by our millimeter observations. Finally, the orange squares display the sources that were confirmed as Cha II cluster member by the analysis of \citet{Galli_2020b}.}}
		\label{fig:comparison}
\end{figure*}

    In Fig.~\ref{fig:comparison}, we display the Herschel 70\,$\mu$m flux of all the Class~II sources observed in \citet{Spezzi_2013}, complemented by the Class I source IRAS12500-7658, and the flat spectrum source J130521.7-773810. We indicate the sources observed by our ALMA observations, with the non-detections marked by blue squares. Finally, we also indicate the sources confirmed as cluster members by \citet{Galli_2020b} using \emph{Gaia DR2} data.

 \section{Cumulative distributions curves}
 \label{sec:cumdistrib_bis}

 \begin{table*}[h!]
 
\caption{Results of pair comparisons with the logrank test applied on the ratio M$_\mathrm{dust}$/M$_\star$ for disks more massive than 1.52 M$_\oplus$. }
\centering
\small
\begin{tabular}{ccccccccc}
\hline \hline
$M_{lim} = 1.52 M_\oplus$ & Upper Sco & {CrA} & Cha II & IC348 & Cha I & Lupus B7 & Lupus B6 & Taurus\\
\hline
Upper Sco&&-& -&-&*&**&*&**\\
CrA&0.9&&-&-&-&*&*&*\\
Cha II&0.5&0.9&&-&-&*&-&*\\
IC348&0.2&0.4&0.7&&-&-&-&*\\
Cha I&0.01&0.1&0.2&0.2&&-&-&-\\
Lupus B7&0.002&0.04&0.04&0.1&0.8&&-&-\\
Lupus B6&0.005&0.08&0.06&0.3&1&0.4&&-\\
Taurus&7e-6&0.05&0.003&0.005&0.1&0.3&0.1&\\
\hline
\end{tabular}
\tablefoot{The lower left half of the table represents p-values for pairs comparisons, and the upper right half corresponds to the interpretation of the test. A p-value lower than 0.001 (**) corresponds to regions statistically distinct, p-values higher than 0.05 (-) to regions statistically similar, and p-values between 0.001 and 0.05 (*) to regions that are marginally different.}
\label{tab:logrank_det}
\end{table*}

\begin{figure}[h!]
	\centering
        \includegraphics[width=9cm]{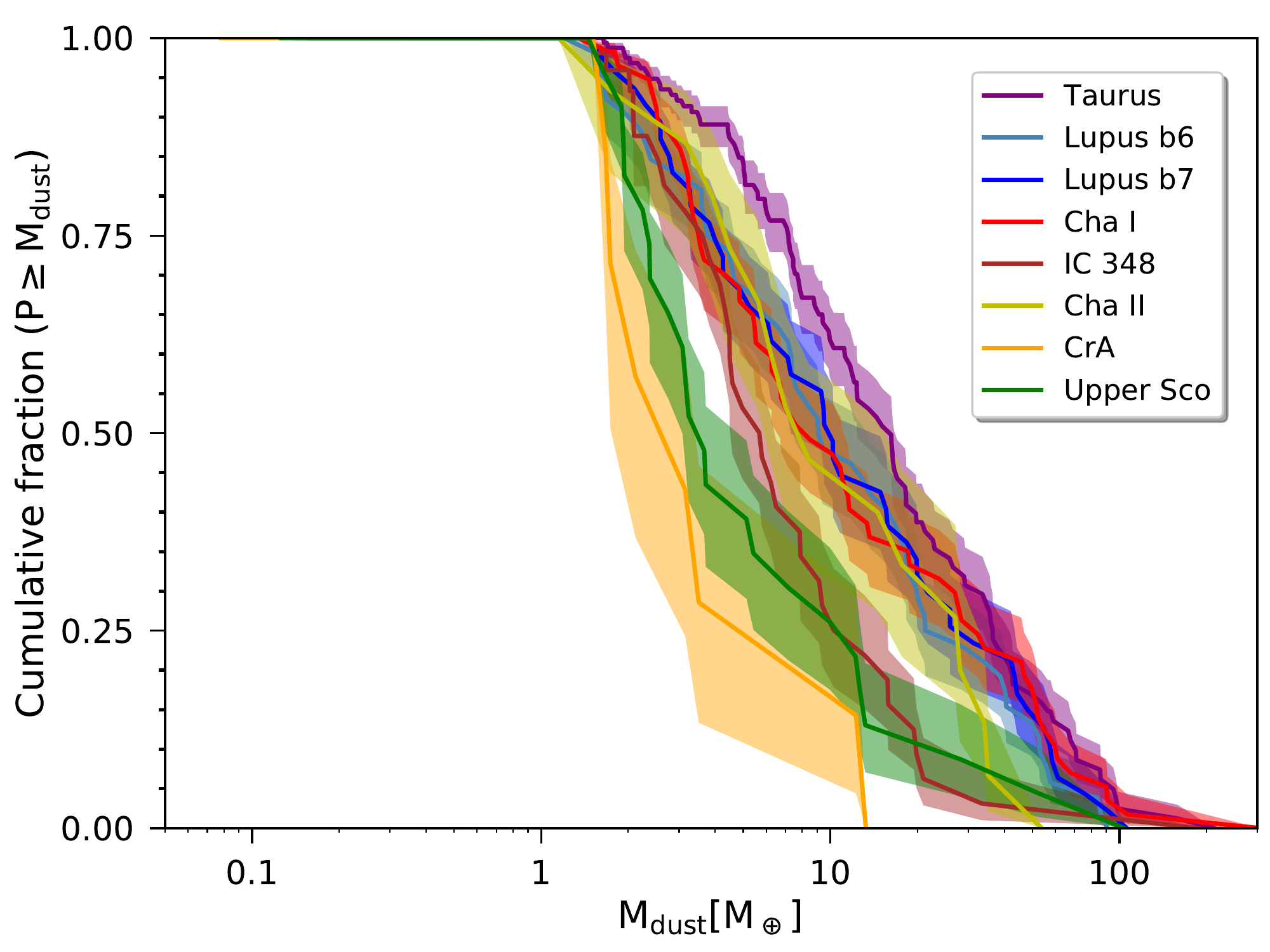}
        \includegraphics[width=9cm]{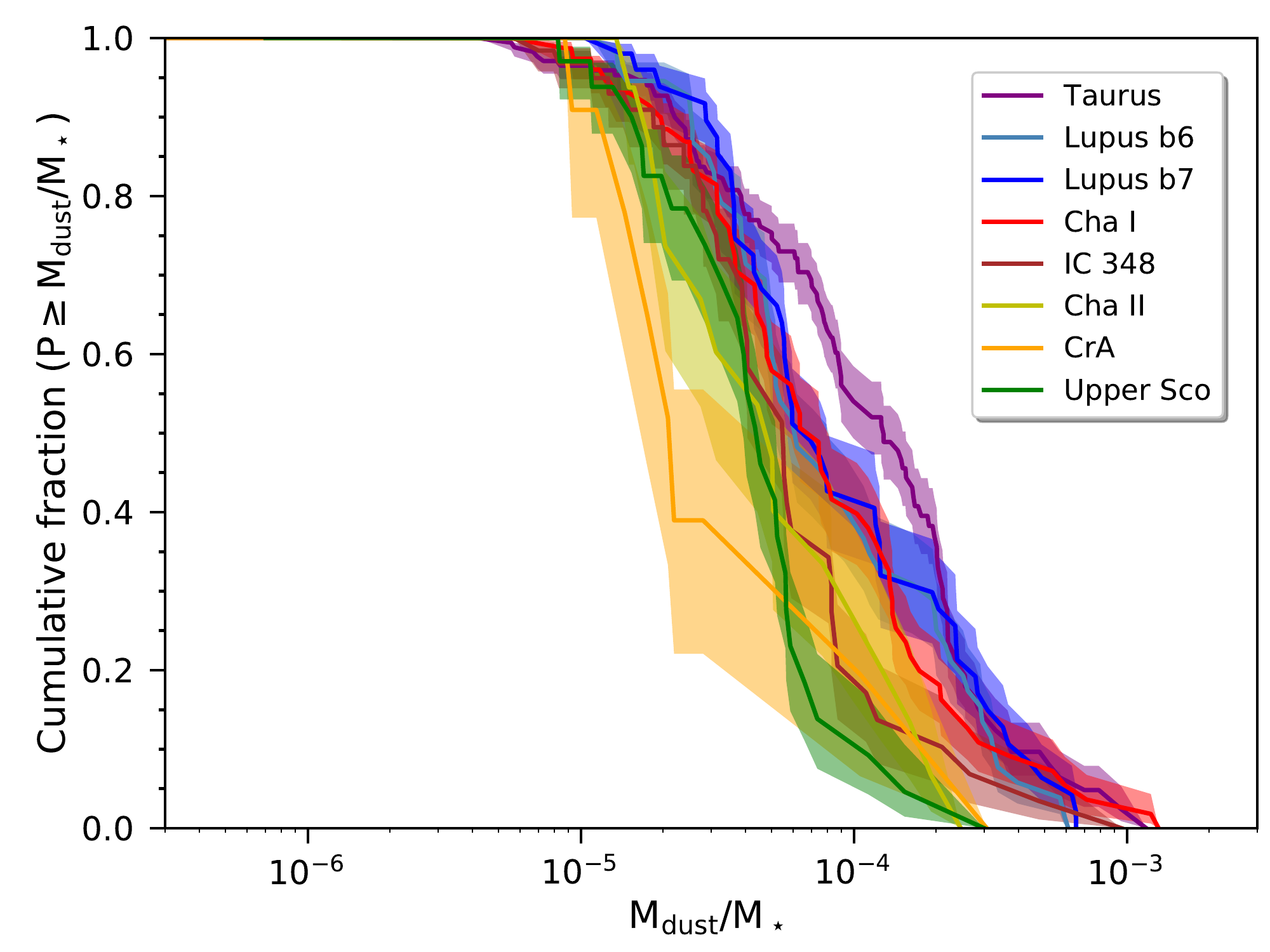}
 		\caption{Cumulative distributions as in Fig.~\ref{fig:distribution_functions} considering disks with smaller masses than $1.52$M$_\oplus$ as non-detections. We normalized the $M_{dust}$ distributions to 1 since the distributions are complete above the $1.52M_\oplus$ limit.} 
	\label{fig:distrib_sameMass}
\end{figure}

Even though the Kaplan-Meier estimator and the logrank test take into account the non-detections, we performed the analysis considering a common dust mass limit. On Fig~\ref{fig:distrib_sameMass}, we show the distributions functions of M$_\mathrm{dust}$ and M$_\mathrm{dust}/$M$_\star$ when considering all systems with disk masses smaller than $1.52$ M$_\oplus$ (the highest M$_\mathrm{dust, min}$ in Table~\ref{tab:regions}) as non-detections. We observe the same ranking in M$_\mathrm{dust}$ as in Sect.~\ref{sec:Luminosity_function}. The results of the logrank test are also presented in Table~\ref{tab:logrank_det}. As expected, they are less statistically different than when the full samples are considered.

\end{appendix}
\end{document}